# Extraordinary linear dynamic range in laser-defined functionalized graphene photodetectors

Adolfo De Sanctis[†], Gareth F. Jones[†], Dominique J. Wehenkel[†], Francisco Bezares[‡], Frank H. L. Koppens[‡], Monica F. Craciun[†], Saverio Russo[†,*]

[†]Centre for Graphene Science, College of Engineering, Mathematics and Physical Sciences, University of Exeter, EX4 4QL Exeter, United Kingdom.

[‡]ICFO - Institut de Ciències Fotòniques, Mediterranean Technology Park, 08860 Castelldefels, Barcelona, Spain.

*To whom correspondence should be addressed, e-mail: S.Russo@exeter.ac.uk

## ABSTRACT

Graphene-based photodetectors have demonstrated mechanical flexibility, large operating bandwidth and broadband spectral response. However, their linear dynamic range (LDR) is limited by graphene's intrinsic hot-carriers dynamic, which causes deviation from a linear photoresponse at low incident powers. At the same time, multiplication of hot carriers causes the photoactive region to be smeared over distances of a few μm, limiting the use of graphene in high-resolution applications. In this work we present a novel method to engineer photoactive junctions in $FeCl_3$-intercalated graphene using laser irradiation. Photocurrent measured at these planar junctions shows an extraordinary linear response with a LDR at least 4500 times larger than other graphene devices (44 dB), while maintaining high stability against environmental contamination without the need for encapsulation. The observed photoresponse is purely photovoltaic, demonstrating complete quenching of hot-carrier effects. These results pave the way towards the design of ultra-thin photodetectors with unprecedented LDR for high definition imaging and sensing.

# INTRODUCTION

Intense research activity on graphene-based photodetectors (*1*) has demonstrated a unique range of properties including mechanical flexibility (*2*), large operating bandwidth (*3*) and broadband spectral response. However, state-of-the-art inorganic (Si, Ga, GaAs, etc.) photodetectors currently exhibit a linear response over a larger range of optical powers as compared to graphene. This is due to the comparatively small density of states in graphene at energies below 1 eV. Furthermore, the thermal diffusion of photo-generated carriers has been found to dominate photocurrent signals measured in graphene-based photodetectors (*4-6*). These strong photothermoelectric effects enable multiplication of hot carriers but also cause photo-responsive regions to be smeared out over distances exceeding 2 microns (*5-7*). The narrow linear dynamic range (LDR) and the size of the photoresponsive regions in graphene photodetectors limits integration of graphene pixels in high resolution sensing and video imaging applications.

Chemical functionalisation (*8*) is a largely unexplored route to overcome the intrinsic limitations on sensing introduced by hot carrier dynamics in pristine graphene, where the limited size of the Fermi surface imposes tight constraints to the carriers relaxation dynamic (*9*). Although attempts have been made to use chemical functionalisation to engineer p-n junctions in graphene (*10,11*) and selectively define photo-responsive regions (*2,12,13*), no major improvements have been shown compared to pristine graphene devices and several challenges remain. These include finding forms of functionalisation which give ultra-high values of charge doping and are also air-stable. Functionalisation of graphene with $FeCl_3$ has been found to result in record high levels of hole-doping ($\approx 1 \times 10^{15}$ cm$^{-2}$) with a room temperature electrical conductivity up to 1000 times larger than pristine graphene whilst maintaining equivalent absorption over the visible wavelength range (*14,15*). At the same time, an unforeseen stability to harsh environmental conditions (*16*), the easy of large-area processing (*15*) and the promise for efficient coupling of telecommunication wavelength light to electrical signals through surface plasmons, make this material uniquely suited to explore novel optoelectronic applications. The development of a new generation of imaging arrays with unprecedented LDR and pixel density, which do not employ any thermal isolation or electrostatic gating at high voltages and are stable in both ambient and harsh conditions, would bring imaging and sensing technologies to new frontiers.

In this work, we demonstrate micro-metre and nano-metre scale planar photo-responsive junctions, which are directly written in the host material using focused laser light. Characterisation of photocurrent signals reveals a purely photovoltaic response and a LDR as large as 44 dB, at least 4500 times larger than any previously reported graphene photodetector (*3,9,17-20*). Crucially, these detectors exhibit remarkable stability in atmospheric conditions without any form of encapsulation and maintain a broad spectral response from UV-A to mid-infrared wavelengths. By employing emerging nano-photonics tools such as near-field photocurrent nanoscopy we are able to surpass the diffraction-limited resolution of far-field methods and define photo-responsive junctions smaller than half the laser wavelength used.

The light-assisted design of integrated and atomically-thin optoelectronic circuits is a step forward to a new frontier in high definition sensing applications, while $FeCl_3$-intercalated few-layer graphene ($FeCl_3$-FLG) defines a new paradigm in ultra-thin, high-LDR photodetectors.

## RESULTS AND DISCUSSION

**Preparation of laser-defined junctions**

The starting material to achieve our goal is an intercalated 4-layer graphene flake with $FeCl_3$ only introduced between the top three carbon layers. Intercalation of $FeCl_3$ molecules into mechanically exfoliated few layer graphene on a $Si/SiO_2$ substrate was conducted using a previously reported method (*14*) in a two zone furnace (see Methods and Materials). A typical Raman spectrum of such a system shows the $G_0$ peak at $1580\ cm^{-1}$ due to the $E_{2g}$ phonon mode of pristine graphene as well as the red-shifted $G_1 = 1615\ cm^{-1}$ and $G_2 = 1625\ cm^{-1}$ peaks of the same mode caused by the charge doping of $FeCl_3$ molecules adjacent to only one side of a graphene layer (stage-2) or sandwiching the carbon atoms (stage-1), see figure 1a. Upon exposure to $532\ nm$ laser light with an incident power of $15.3\ MW/cm^2$ for $3\ s$, we observe a drastic modification of the Raman G-band: with a pronounced down-shift of the G-peak positions; a reduction of their full width at half maximum (FWHM) and the disappearance of the $G_2$ peak and the emergence of the $G_0$ peak (see figure 1a). All of these changes indicate a reduction in hole doping caused by laser-induced displacement of $FeCl_3$, with the disappearance of the $G_2$ peak stemming from the complete removal of stage-1 intercalation. Finally, the absence of a defect-related Raman peak demonstrates that this functionalisation can truly sustain laser powers more than 300 times higher than pristine graphene (Supplementary Information S1).

To ascertain the effectiveness of laser irradiation as a method for locally tailoring $FeCl_3$ intercalation in graphene, we exposed a $5.5~\mu m$ wide section of the intercalated flake to a raster laser scan ($15.3~MW/cm^2$ for $3~s$ in $0.5~\mu m$ steps). Raman spectra were collected at incrementally spaced locations across the laser-exposed region both before and after illumination, as shown in figure 1b. Comparing the spectral profiles at each location, it is apparent that all irradiated regions undergo a substantial degree of de-intercalation. In figure 1c, we quantify changes in chemical structure across the entire laser-exposed region by analysing the positions of the $G_1$ and $G_2$ peaks along a $21~\mu m$ line scan. Uniform removal of the $G_2$ peak from the entirety of the rastered region clearly demonstrates that $FeCl_3$ molecules may be displaced from arbitrarily mapped areas. Importantly, the degree of intercalation remains unchanged away from the irradiated area, with the resolution of $FeCl_3$ displacement defined by the laser spot profile. The remarkable effectiveness of laser-induced de-intercalation over a significant fraction of the $FeCl_3$-FLG flake area presents an elegant method, akin to optical lithography, which can be used to locally customise the chemical functionalisation of graphene layers.

The shift of the Raman G-peak is quantitatively translated into a charge density using the model developed by Lazzeri *et al.* (*21*) and Das *et al.* (*22*) with an accuracy of $\pm 10\%$ as shown by independent characterization of charge density from quantum oscillations in magnetoconductance (*14,15*). We find that the laser irradiation of $FeCl_3$ causes a reduction in charge density of up to $\Delta p_{tot} \approx -0.6 \times 10^{14}~cm^{-2}$ (figure 2a) which agrees well with electrical measurements showing a 170% increase in resistivity over the modified area (see Supplementary Information S4). Hence, the abrupt change in hole concentration at the boundaries of the laser-exposed region defines sharp p-p' junctions (see Supplementary Information S2.5 for data on additional devices).

**Optoelectronic response of laser-defined p-p' junctions**

Inspired by the rich variety of charge transfer processes which has enabled a revolution in semiconductor heterostructures applications, we examined the optoelectronic response of these laser-defined junctions in FeCl$_3$-FLG. Laser light focused to a beam spot diameter of 1.0 μm at 300 μW was rastered over the device surface whilst measuring photocurrent signals, see figure 2b. Photocurrent maps are given in figure 2c for a variety of excitation wavelengths. The sign convention of the photocurrent has been carefully configured so that a positive signal indicates the drift of holes from the left to the right electrode (Supplementary Information S5.4). As expected for uniform doping, no significant photocurrent is observed in FeCl$_3$-FLG before laser patterning. However, when a p-p'-p junction is defined by laser-assisted displacement of FeCl$_3$, a photocurrent as large as 9 nA is measured at each of the lateral interfaces.

A multitude of physical mechanisms can give rise to a photoresponse. Of these, two play a major role in graphene-based photodetectors. They are the photothermoelectric (PTE) and the photovoltaic (PV) effect (*1*). The PTE originates from a difference in Seebeck coefficients, $\Delta S = (S' - S)$, across a graphene junction formed by regions with a differing density of states. If the junction is illuminated, a local increase of temperature ($\Delta T$) results in the diffusion of carriers and an opposing photovoltage ($V_{PTE} = \Delta S\, \Delta T$) is generated. Hot carrier dynamics are generally recognized to dominate photocurrent generation in supported graphene devices due to inefficient cooling of electrons with the lattice (*5,6*). For the PV effect, incident photons generate a density ($n_{ph}$) of carriers which, in the presence of an in-built electric field, are separated and induce current at the electrodes (figure 2b). Other mechanisms such as the bolometric effect, photogating effect and Dyakonov-Shur effect require an externally applied voltage (*1*) and are therefore not active in the short circuit configuration of our measurements (figure 2b).

A first insight on the microscopic mechanism behind the observed photocurrent can be gained by comparing the laser power dependence in pristine and intercalated graphene. Figure 3a shows a typical power dependence for photocurrent ($I_{PH} \propto P^{\alpha}$) generated in one of several measured monolayer graphene devices (Supplementary Information S2.4) where $\alpha = 2/3$ was obtained with $10~\mathrm{mV}$ applied between source and drain. On the other hand, the photoresponse in FeCl$_3$-FLG is strikingly different from that of pristine graphene, exhibiting a linear dependence extending beyond three logarithmic decades of incident laser power. The observed difference originates from the charge carrier dynamics. More specifically, in pristine graphene the chemical potential ($\mu$) lies close to the charge neutrality point and the small Fermi surface imposes tight constraints on the maximum energy lost through momentum-conserving acoustic phonon emission ($\Delta E_{ac} < 2\hbar v_s k$, where $v_s \sim 2 \times 10^4~\mathrm{ms}^{-1}$ is the acoustic phonon speed and $k$ is the hot carrier wavenumber) (23). As a result, photo-excited carriers reach a steady state temperature far above that of the lattice ($T_h \gg T_l$) and are instead cooled via short-range "supercollision" processes at sites of disorder (9,24). If the PTE effect is similarly responsible for photocurrent in FeCl$_3$-FLG, the steady state temperature of hot carriers must lie significantly closer to that of the lattice ($T_h - T_l \ll T_l$) in order to justify the observed linear power dependence (9). A reduction in $T_h$ can be explained by the ultrahigh levels of charge density (up to $3 \times 10^{14}~\mathrm{cm}^{-2}$ per layer) achieved through FeCl$_3$ intercalation (14); the expanded Fermi surface enhances $\Delta E_{ac}$ to as much as $60$ times that of pristine graphene, accelerating the cooling of photo-generated charges. On the other hand, the small temperature gradients present at these highly doped junctions could diminish thermoelectric currents so much that they become negligible compared to signals generated by the PV effect. A linear power dependence would also be expected in this case (25), provided that the incident light intensity is sufficiently low so as to not affect the average lifetime ($\tau$) of photo-generated carriers. The observation of photocurrent with a linear dependence upon incident power therefore indicates enhanced cooling of hot carriers in FeCl$_3$-FLG but cannot, as other studies have suggested (19), be used independently to distinguish between PTE and PV effects.

**Photovoltaic effect in FeCl$_3$-FLG junctions**

In order to identify the origin of photocurrent at p-p' junctions of FeCl$_3$-FLG, we adapt the model of Song *et al.* (5) to calculate the relative contributions of the PTE and PV effects (Supplementary Information S5). The photocurrent produced in a p-p' junction located in the middle of an FeCl3-FLG channel (length $L$ and width $W$) illuminated by a laser (spot diameter $l_0$) is:

$$I_{ph} = \int_0^W \int_{-\frac{L}{2}}^{\frac{L}{2}} [S_{(x,y)} \nabla T_{(x,y)} - \sigma_{(x,y)}^{-1} n_{ph\,(x,y)} \eta \nabla \mu_{(x,y)}] \frac{dydx}{RW}, \quad (1)$$

where $R$ is the resistance of the graphene layer and $\eta$ the carrier mobility. For a doped graphene layer with a charge carrier density above $n \approx 3 \times 10^{13}$ cm$^{-2}$, the Bloch-Grüneisen temperature ($T_{BG} = \Delta E_{ac}/k_B$) exceeds $300$ K (*26*). Therefore, under continuous wave illumination, where $\Delta T$ is typically just a few Kelvin (*7*), bottlenecks in electron-acoustic phonon coupling are alleviated in FeCl3-FLG. The increased efficiency of momentum-conserving acoustic phonon emission renders supercollisions irrelevant to hot carrier cooling processes and reduces the average cooling length ($\zeta$) from several microns (*5,7*) to approximately $200$ nm. Hence, for a typical device $\zeta \ll L/2$. Using the low energy density of states for monolayer graphene and a minimum conductivity of $\sigma_{min} \approx 4e^2/h$ (*27*), we express the conductivity of each decoupled layer as a function of its chemical potential $\sigma(\mu) = \sigma_{min}(1 + \mu^2/\Lambda^2)$ where $\Lambda \approx 140$ meV. The Mott relation for thermopower (*27*) and a solution to the heat equation which assumes non-divergent current densities (*5*) are then used with equation (1) to estimate the relative magnitudes of PTE and PV currents from the electrical properties either side of the p-p' junction:

$$\frac{I_{PTE}}{I_{PV}} = \frac{2e\,k_B T_h l_0 \sigma_{min}}{\mu \mu' \eta \tau \Lambda} \cdot \frac{\left[\mu'\left(1 - \frac{\sigma_{min}}{\sigma}\right) - \mu\left(1 - \frac{\sigma_{min}}{\sigma'}\right)\right]}{\left(\frac{\sigma}{\zeta} + \frac{\sigma'}{\zeta'}\right) \cdot \left[\tan^{-1}\left(\frac{\mu}{\Lambda}\right) - \tan^{-1}\left(\frac{\mu'}{\Lambda}\right)\right]}, \quad (2)$$

with $1$ ps $< \tau < 2$ ps in good agreement with pump-probe spectroscopy measurements (*28*) (see Supplementary Information S5), and all material parameters are averaged over the device width. For each of the decoupled monolayers in the four layer flake, where the most prominent changes in chemical potential occur after laser writing, we calculate $I_{PTE}/I_{PVE} \approx -0.06$ using equation (2). Thermalisation of hot carriers therefore makes a negligible contribution to the total photocurrent generated at FeCl3-FLG p-p' junctions and acts in the opposite direction to dominant photovoltaic processes. Opposing photocurrents at p-p' junctions have previously been predicted in monolayer graphene transistors with split electrostatic gates (*5*) and can be understood intuitively by considering that the movement of photo-generated charge carriers is governed by local gradients in chemical potential for photovoltaic currents and by local gradients in Seebeck coefficient in the case of thermoelectric currents. Following the Mott relation ($S \propto -\sigma^{-1}(d\sigma/d\mu)$), the density of states of graphene dictates that these gradients will always point in opposite directions so long as the chemical potentials each side of a photo-active junction are both situated in the valence band (p-p' junctions) or both in the conduction band (n-n' junctions) away from the charge neutrality point. As a result of these findings, we take the direction of photocurrent signals shown in figure 2c (where carriers drift according to the local potential gradient at p-p' interfaces) as direct evidence of a purely photovoltaic response in laser-written FeCl$_3$-FLG detectors.

The selective quenching of thermoelectric processes in graphene through chemical functionalisation could prove to be a highly useful tool for extending the use of graphene-based light sensors beyond micro-bolometers and modulators suitable for infra-red wavelengths. Pixels of FeCl$_3$-FLG-based photodetectors would not require thermal isolation and could be packed to a far higher density than undoped graphene monolayers, making them well-suited for imaging applications over a broad spectral range.

**Extraordinary linear dynamic range**

The purely PV response in FeCl$_3$-FLG detectors is characterized by an extraordinary LDR. The noise-equivalent-power (NEP) of our device was measured to be $4\ kW/\text{cm}^2$ (see Supplementary Section S2.2), thus resulting in a LDR of $44\ \text{dB}$. This is $4500$ times larger than previously reported graphene photodetectors (LDR $\approx 7.5\ \text{dB}$) (*3*) and $\sim 800$ times larger than other functionalized graphene devices (LDR $\approx 15\ \text{dB}$) (*13*). In Supplementary Table S1 we show a comparison of the maximum saturation power and LDR for different devices reported in literature (see also Supplementary Section S2.3 for a comparative study of detectors).

In order to further asses the suitability of FeCl$_3$-FLG for optoelectronic applications, we have characterised the photoresponse at these p-p' junctions over a wide range of light intensities and wavelengths. Figure 3b shows the power dependence of photocurrent measured at a p-p' junction in FeCl$_3$-FLG for various wavelengths of incident light ranging from UV-A (375 nm) to red (685 nm). Fits of the power exponent at each wavelength give: $\alpha_{375} = 0.99 \pm 0.01$, $\alpha_{473} = 1.05 \pm 0.06$, $\alpha_{514} = 0.97 \pm 0.03$, $\alpha_{561} = 0.99 \pm 0.01$ and $\alpha_{685} = 0.95 \pm 0.05$. For the multitude of FeCl$_3$-FLG devices measured, we observed no deviation from a strictly linear power dependence in the whole measured power range. This indicates that the ultra-high degree of charge carrier doping introduced by FeCl$_3$ intercalation acts as a uniquely stable method to quench thermoelectric effects and fix the photoresponse to an extended linear dynamic regime, avoiding the sensitivity to processing methods and environmental conditions which pristine graphene photodetectors (*3, 9*) inevitably suffer from. In figure 3c, the spectral responsivity, $\Re(\lambda) = I_{ph}/P_{opt}(\lambda)$, of a p-p' junction is displayed with and without correcting for reflections from the Si/SiO$_2$ substrate (Supplementary Information S6). The photoresponse remains remarkably consistent across the entirety of the visible range, where $\Re(\lambda)$ varies by only one order of magnitude, with values $> 0.1$ mA/W, which are typical for high-end all-graphene photodetectors (*1*). Of particular interest is the increase in responsivity towards UV-A wavelengths, a region where the performance of silicon photodiodes decreases. We attribute the extended LDR to accelerated carrier cooling and the enhanced responsivity to an increased high energy density of states introduced by FeCl$_3$-intercalation of graphene (*28*). This consistent proportionality between output electrical signal and incident optical power over a broad spectral range makes FeCl$_3$-FLG-based photodetectors ideally suited to radiometry and spectroscopy applications.

**Below the diffraction-limit**

The spatial resolution of FeCl$_3$ displacement at the engineered p-p' junctions is determined by the profile of the laser spot used for patterning. In far-field optical microscopy, spot sizes are dictated by the Abbe diffraction-limit ($\sim \lambda/(2NA)$, where $NA$ is the numerical aperture of the objective). In order to explore the density to which graphene-based imaging pixels may be packed in the absence of hot carrier effects, we employ scattering-type near-field optical microscopy (s-SNOM, see Methods) to define photo-active junctions below the Abbe limit. This technique has been used extensively to study the plasmonic (*29*) and optoelectronic (*30*) response of graphene-based devices. Figures 4a-c show photocurrent maps, using a $\lambda = 10$ μm excitation source, taken before and after displacement of by a $\lambda = 632$ nm laser. Planar junctions exhibiting a photovoltaic response are readily defined with a peak-to-peak separation of just $250$ nm (figure 4f) whilst concurrent topography mapping (figures 4d-e) indicates that the flake surface remains undamaged. Furthermore, the photocurrent is stronger near the edges of the flake, suggesting that the de-intercalation process is due to the displacement of FeCl$_3$ molecules in the plane of graphene which are removed from the edges. The absorption of photons with energy $E << 2\mu$ in FeCl$_3$-FLG highlights the role of transitions to the $\pi$ band from localized states introduced by FeCl$_3$, as predicted by DFT calculations (*31*). This prevents Pauli blocking of long wavelengths and maintains a broadband spectral response, up to mid-infrared (MIR) wavelengths, in these novel photodetectors.

## CONCLUSIONS

In conclusion, laser-patterning is an elegant method of creating photo-responsive junctions in intercalated few-layer graphene. Photo-responsive junctions in FeCl$_3$-FLG are engineered on the sub-micron scale and remain highly stable under atmospheric conditions and intense light exposure. This presents a unique opportunity relative to other methods of chemical functionalisation, whereby photocurrent mechanisms are reliably pinned to produce a linear response over broad ranges of power and wavelength with no requirement for encapsulation from the environment. These junctions show an extraordinary linear dynamic range up to $44$ dB, more than $4500$ times larger than other graphene photodetectors, that can operate at incident optical powers up to $10^4$ kW/cm$^2$ in the whole visible range, in the near-UV and at MIR wavelengths. Further enhancements to responsivity can be achieved through the use of an increased number of intercalated graphene layers and optimisation of the de-intercalation process to maximise the chemical potential gradient at p-p' junctions. Uniform intercalation of FeCl$_3$ throughout large-area graphene films of a uniform layer number will be crucial for implementing these findings in practical applications. To this end, intercalation of large-area CVD-grown graphene has already been demonstrated (*15,32,33*) and roll-to-roll processing of graphene is readily applicable to intercalated films. Compact pixels arrays could be realized using vertical circuitry equivalent to buried channels in CMOS technology, where vias connect between pixels on the substrate surface and laterally running interconnects dispersed over several buried levels. These findings provide exciting prospects for light detection in laser-induced plasmas; UV photocatalytic water sanitation processes; and high precision manufacturing. In such environments, these novel sensors could eliminate the need for attenuating optics in the detection of ultra-bright light signals with high spatial resolution.

## MATERIALS AND METHODS

**Device fabrication.** Few layer graphene flakes were mechanically exfoliated from natural graphite on a p-doped Silicon substrate with a $280\,\mathrm{nm}$ surface oxide. Intercalation with $FeCl_3$ was conducted in a two-zone furnace using a previously demonstrated vapour transport method (*14*). Electrical contacts to the flakes were defined by standard electron-beam lithography, thermal deposition of Cr/Au ($5/50\,\mathrm{nm}$) and lift-off in acetone.

**Raman spectroscopy.** Raman spectroscopy measurements used to characterise the degree of intercalation in $FeCl_3$-FLG were performed in atmosphere and at room temperature (see Supplementary Information). Raman spectra were acquired with a Renishaw spectrometer equipped with a $532\,\mathrm{nm}$ laser focused to a $1.0\,\mathrm{\mu m}$ spot through a 50× objective lens. An incident power of $1\,\mathrm{mW}$ was used for all measurements and spectra were recorded with a $2400\,\mathrm{g/mm}$ grating. A CCD acquisition time of 5 seconds was used.

**Photocurrent measurements.** A continuous wave laser beam from a multi-wavelength ($375\,\mathrm{nm}$, $473\,\mathrm{nm}$, $514\,\mathrm{nm}$, $561\,\mathrm{nm}$, $685\,\mathrm{nm}$) solid-state laser diode array was focused onto the sample through a 50× lens, producing a spot-size of $1.0\,\mathrm{\mu m}$. A high resolution microscope stage (min step-size of $0.05\,\mathrm{\mu m}$ was used to produce spatial maps of the photocurrent. Electrical measurements were performed in short-circuit (zero-bias) configuration using a DL Instruments Model 1211 current amplifier connected to a Signal Recovery model 7124 DSP lock-in amplifier. The lasers were modulated at a frequency of $73.3\,\mathrm{Hz}$ with a TTL signal from a DDS function generator which was used as a reference signal for the lock-in. All measurements were performed at ambient conditions ($T = 300\,\mathrm{K}$, $P = 1\,\mathrm{atm}$) in air. The laser power was varied from $1.5\,\mathrm{\mu W}$ to $1\,\mathrm{mW}$ by means of analog modulation of the laser diodes and the use of neutral density filters (ND) along the beam path. All the devices studied have been measured in air over a time scale longer than 1 year, during which no change in the photoresponse was observed.

**LDR calculation.** The linear dynamic range (LDR) is defined as:

$$LDR = 10 \times log_{10}\left(\frac{P_{sat}}{NEP}\right) [dB], \quad (3)$$

where the Noise Equivalent Power (NEP) is defined as the power at which the signal to noise ratio (SNR) has a value of 1. The NEP can be measured directly or computed as $\text{NEP} = S_I/\Re \; [\text{W}/\sqrt{\text{Hz}}]$, where $S_I$ is the rms current noise (in $\text{A}/\sqrt{\text{Hz}}$) and $\Re$ is the responsivity of the photodetector (in $\text{A}/\text{W}$).

**s-SNOM measurements** Scattering-type Near Filed Optical Microscopy (s-SNOM) involves focusing a laser onto a metallised AFM tip which creates a strong, exponentially-decaying field at its apex. The tip is then scanned across the sample, operating in tapping mode, allowing parameters including topography and scattered light emission to be measured with sub-wavelength resolution (*35-37*). If the device is contacted as in this work, the local photo-current, produced by the light focused at the tip, can be measured with the same resolution. s-SNOM measurements were performed using a commercially available system from Neaspec GmbH. The AFM tips used were commercially available metal-coated tips with average radii of $25 \text{ nm}$. Our system was equipped with a tunable $CO_2$ laser as well as a visible wavelength HeNe laser. In this experiment, the $CO_2$ laser was used to probe the optical near-field signal of our samples, while the visible laser was used only for laser patterning of the p-p' junctions in our devices. Concurrent photocurrent and AFM topography measurements were performed in short-circuit configuration using the $CO_2$ laser before and after laser patterning.

## SUPPLEMENTARY MATERIALS

Supplementary information for this work is submitted in conjunction with the main manuscript:

Section S1 (figures S1 and S2), supplementary data on laser irradiation;
Section S2 (figures S3-S7, and table S1), supplementary photocurrent measurements;

Section S3 (table S2), power dependence of the photothermoelectric and photovoltaic effects;

Section S4 (figure S8), estimation of chemical potential and conductivity for decoupled graphene layers;

Section S5 (figure S9), physical explanation for a purely photovoltaic response;

Section S6 (figure S10, table S3), correction of responsivity spectra for substrate reflections.

References *38-47*.

## ACKNOWLEDGMENTS

**Funding:** S. Russo and M.F. Craciun acknowledge financial support from EPSRC (Grant no. EP/J000396/1, EP/K017160/1, EP/K010050/1, EP/G036101/1, EP/M001024/1, EP/M002438/1), from Royal Society international Exchanges Scheme 2012/R3 and 2013/R2 and from European Commission (FP7-ICT-2013-613024-GRASP). **Author contributions:** D.J.W. conceived the initial experiment. A.D.S. prepared the intercalated graphene, fabricated the devices, performed the Raman and photo-current characterization and analysed the data. G.F.J. assisted with initial measurements and performed the analysis and interpretation of the photo-current response with key contributions by F.H.L.K. and F.B. The s-SNOM measurements were performed by F.B. with the assistance of A.D.S. A.D.S. and G.F.J. wrote the manuscript with contributions and ideas from all the authors. M.F.C. and S.R. supervised the whole project. **Competing interests:** The authors declare that they have no competing interests. **Data and materials availability:** All data needed to evaluate the conclusions in the paper are present in the main text and in the Supplementary Materials. Additional data available from the authors upon request. Correspondence and requests for materials should be addressed to S. Russo.

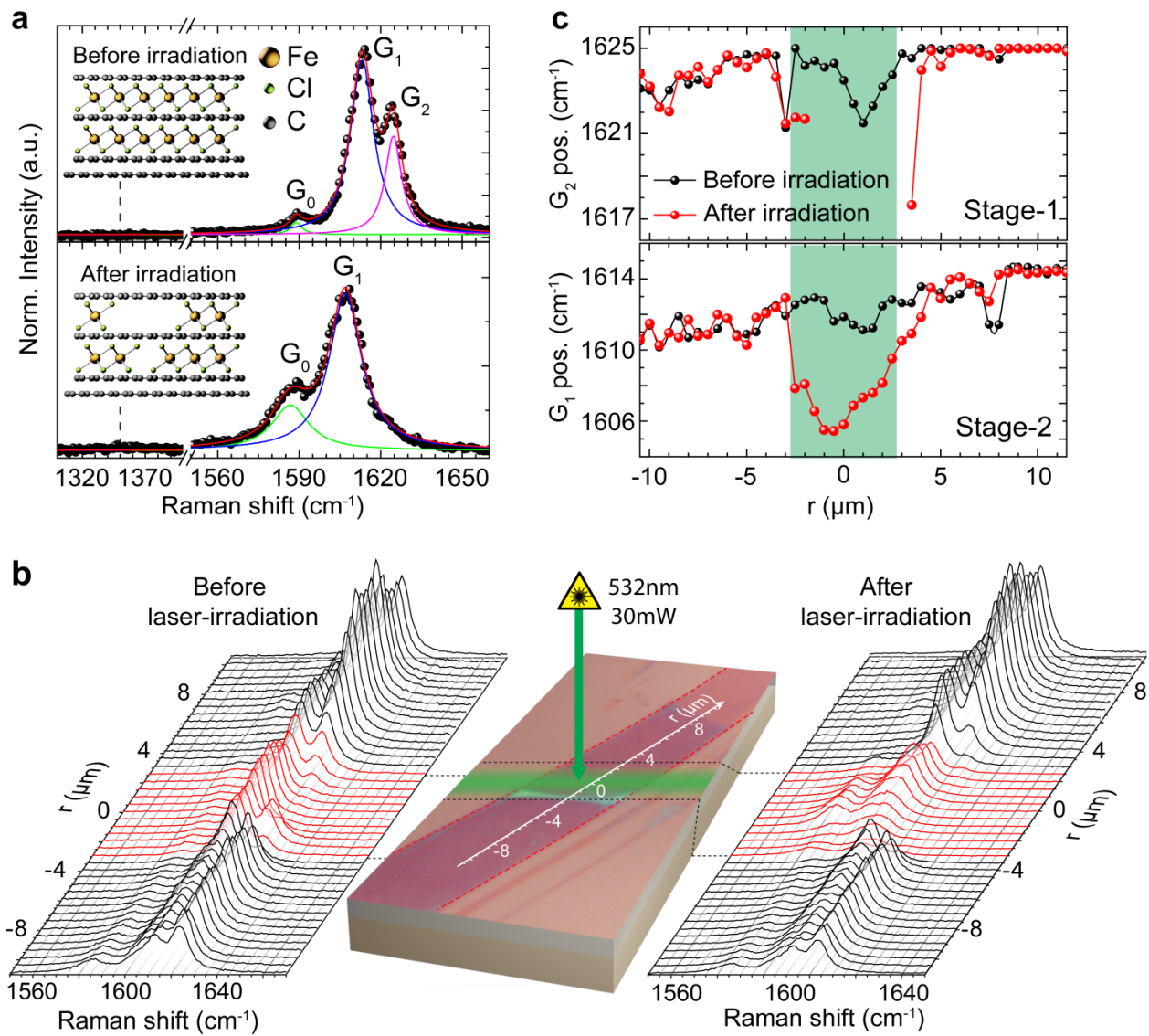

**Fig. 1. Raman spectroscopy study of structural changes in laser-irradiated FeCl$_3$-FLG.** (**a**) G-bands in FeCl$_3$-FLG before (top) and after (bottom) exposure to a $30\,mW$ laser for $3\,s$ ($\lambda = 532\,nm$). Experimental data (black dots) is shown alongside a superposition of Lorentzian fits to the G$_0$, G$_1$ and G$_2$ peaks (red line). (**b**) Optical micrograph of the FeCl$_3$-FLG flake (red-dotted lines) with the laser-irradiated region highlighted (green). Raman spectra are acquired along $r$ before (left) and after (right) FeCl$_3$ displacement. (**c**) G$_1$ (bottom) and G$_2$ (top) peak positions representing stage-1 and stage-2 intercalated states respectively. Data points are Lorentzian fits of the spectral peaks in (**b**).

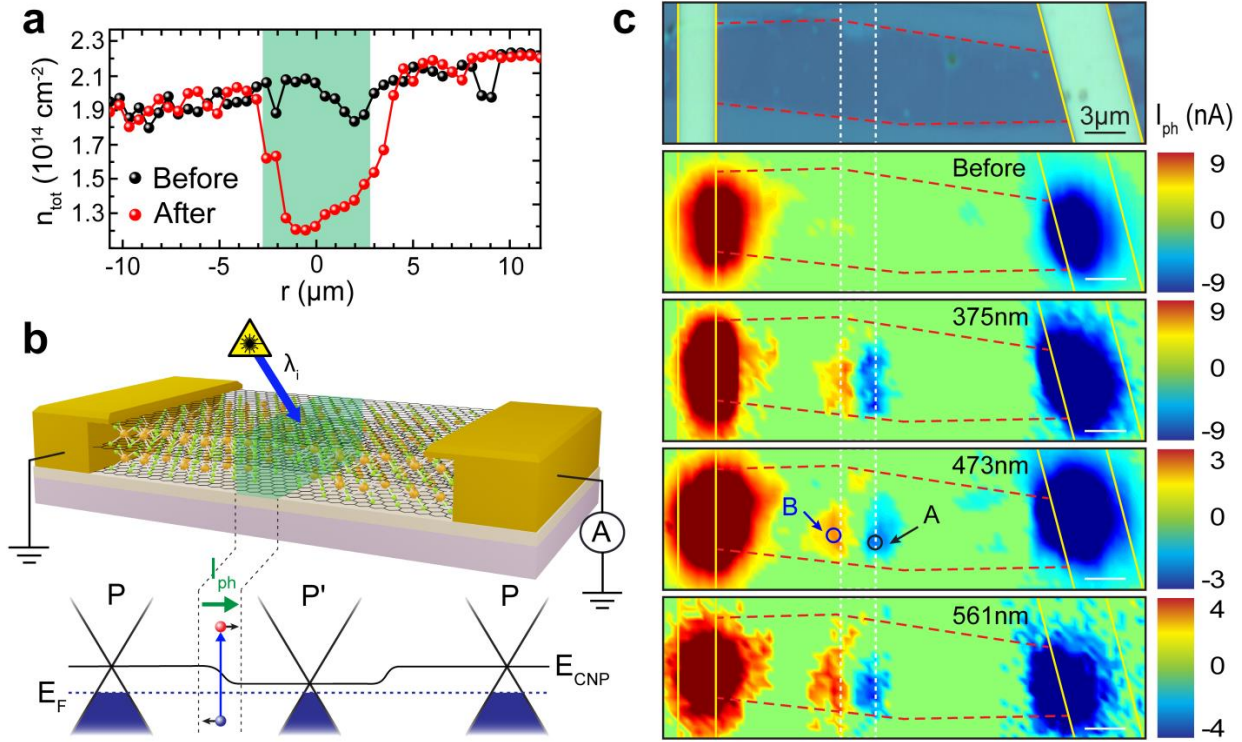

**Fig. 2. Scanning photocurrent microscopy of p-p' junctions in -FLG.** (**a**) Total charge carrier concentration before and after laser-assisted displacement of FeCl$_3$, estimated from G-peak positions in figure 1c. (**b**) Short-circuit configuration (top) for scanning photocurrent measurements of a p-p'-p junction in (p' region in green). Schematic band structure (bottom) of each region illustrates of photo-generated carriers drifting under a chemical potential gradient. (**c**) Optical micrograph (top) of a FeCl$_3$-FLG flake (red-dashed lines) with Au contacts (yellow lines). Scanning photo-current maps (bottom panels) before and after selective laser-assisted displacement of (white-dashed lines). The photoresponse is measured for excitation wavelengths of $375\ nm$, $473\ nm$ and $561\ nm$.

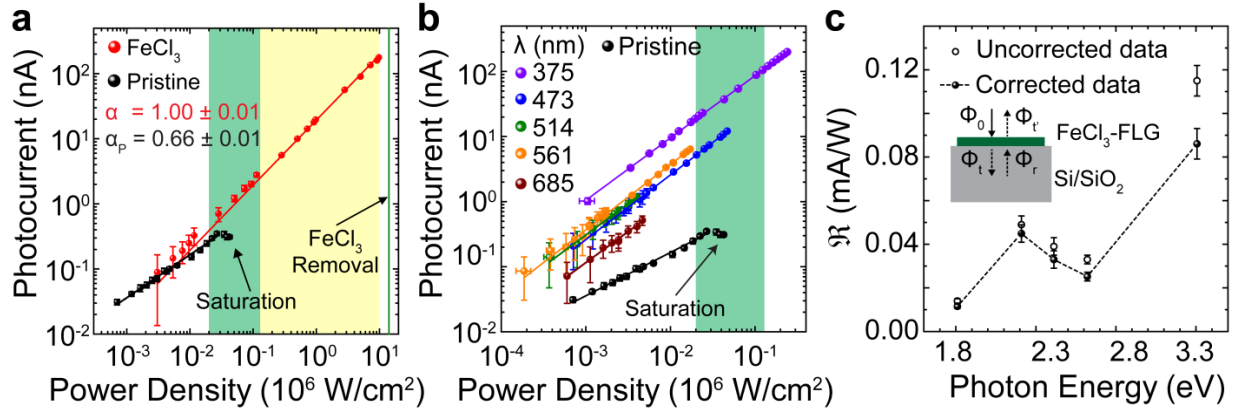

**Fig. 3. Characterisation of photocurrent at p-p' junctions in FeCl$_3$-FLG.** (**a**) Photocurrent produced by $\lambda = 473\ nm$ excitation as a function of incident power density measured at a laser-defined p-p' junction and for pristine monolayer graphene (black). Power-law exponents ($I_{ph} \propto P^\alpha$) are detailed for each data set with fits shown as solid lines. Powers within the range at which photocurrent in pristine graphene has been reported to saturate are highlighted in green (see Supplementary Table S1). Yellow-shaded area represents the extended range of FeCl$_3$-FLG. (**b**) Photocurrent measured at the p-p' junction A in figure 2b using various excitation wavelengths, solid lines are linear fits (see main text). (**c**) Spectral responsivity of a p-p' junction in FeCl$_3$-FLG shown with (filled circles) and without (open circles) correcting for reflections from the Si/SiO$_2$ substrate (Supplementary Information S6), extrapolated from panel (**b**). Dashed line is a guide to the eye. Inset: schematic of the model used to correct $\Re(\lambda)$ for substrate reflections. Power density and responsivity values are calculated considering the area illuminated by the laser spot (see Methods).

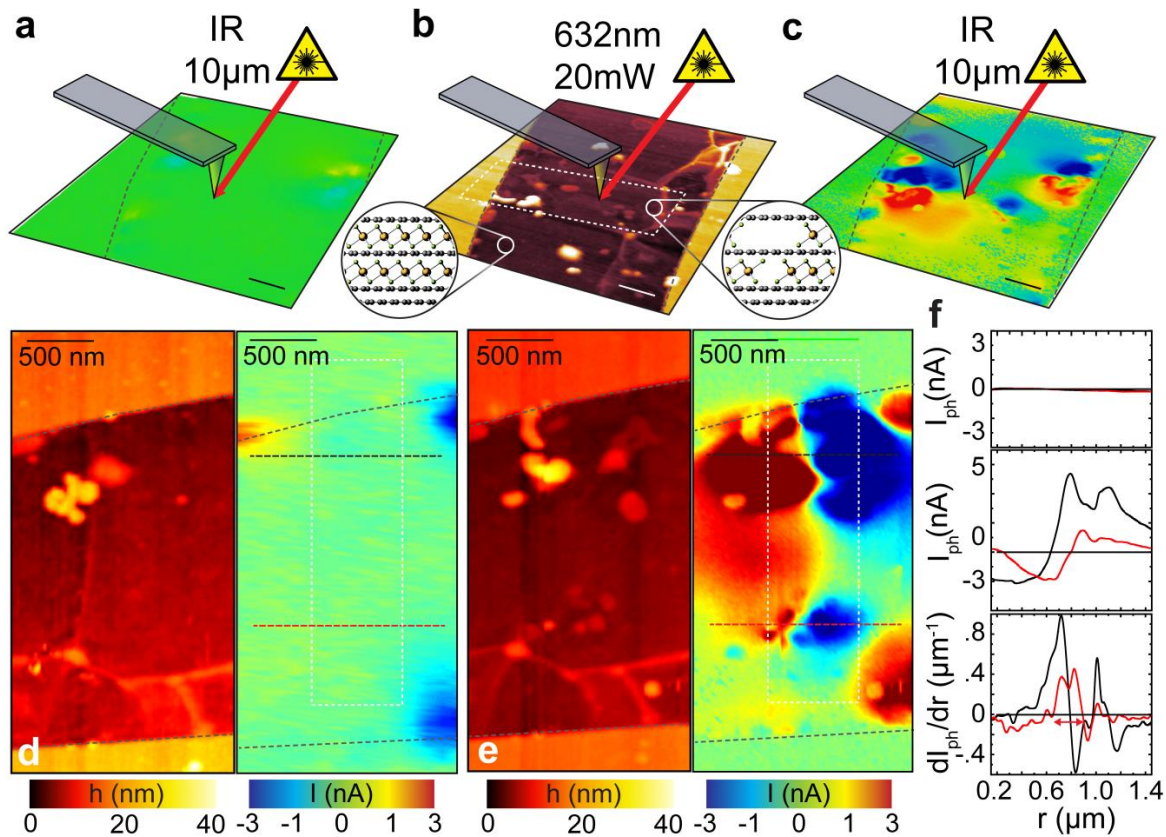

**Fig. 4. High resolution photo-active junctions in -FLG defined using near-field scanning microscopy.** (**a**) Spatial map of photocurrent in a uniformly-doped -flake before laser-assisted de-intercalation. (**b**) AFM topography and (**c**) scanning photocurrent maps of the FeCl$_3$-FLG flake after laser-assisted de-intercalation by a $\lambda = 632\ nm$ laser scanned over a $500\ nm$ long region (white dashed lines). Insets: illustrations of the chemical structure in p- and p'-doped regions. Schematic of the excitation wavelength focused on a metallized AFM tip in each measurement are included in (**a**)-(**c**), outlines of the flake are superimposed (black dashed lines). Scale bars, $500\ nm$. Magnified concurrent AFM topography and scanning photocurrent maps are shown before, (**d**), and after, (**e**), laser writing. (**f**) Line scans of photocurrent measured cross laser-defined p-p'-p junctions ((**d**) and (**e**), red and black dashed lines) before (top panel) and after (middle panel) displacement of molecules. First derivative plots of the photocurrent signal after displacement (bottom panel) shows a peak-to-peak distance of $250\ nm$ between adjacent p-p' junctions (red arrows). All photocurrent measurements were taken in short circuit configuration.

# SUPPLEMENTARY MATERIAL
# Extraordinary linear dynamic range in laser-defined functionalized graphene photodetectors

Adolfo De Sanctis[†], Gareth F. Jones[†], Dominique J. Wehenkel[†], Francisco Bezares[‡],
Frank H. L. Koppens[‡], Monica F. Craciun[†], Saverio Russo[†,*]

[†]Centre for Graphene Science, College of Engineering, Mathematics and Physical Sciences, University of Exeter, EX4 4QL Exeter, United Kingdom. [‡]ICFO - Institut de Ciències Fotòniques, Mediterranean Technology Park, 08860 Castelldefels, Barcelona, Spain.

*To whom correspondence should be addressed, e-mail: S.Russo@exeter.ac.uk

## CONTENTS

# S1 Supplementary data on laser irradiation

## S1.1 Determination of the stacking order in FeCl$_3$-FLG

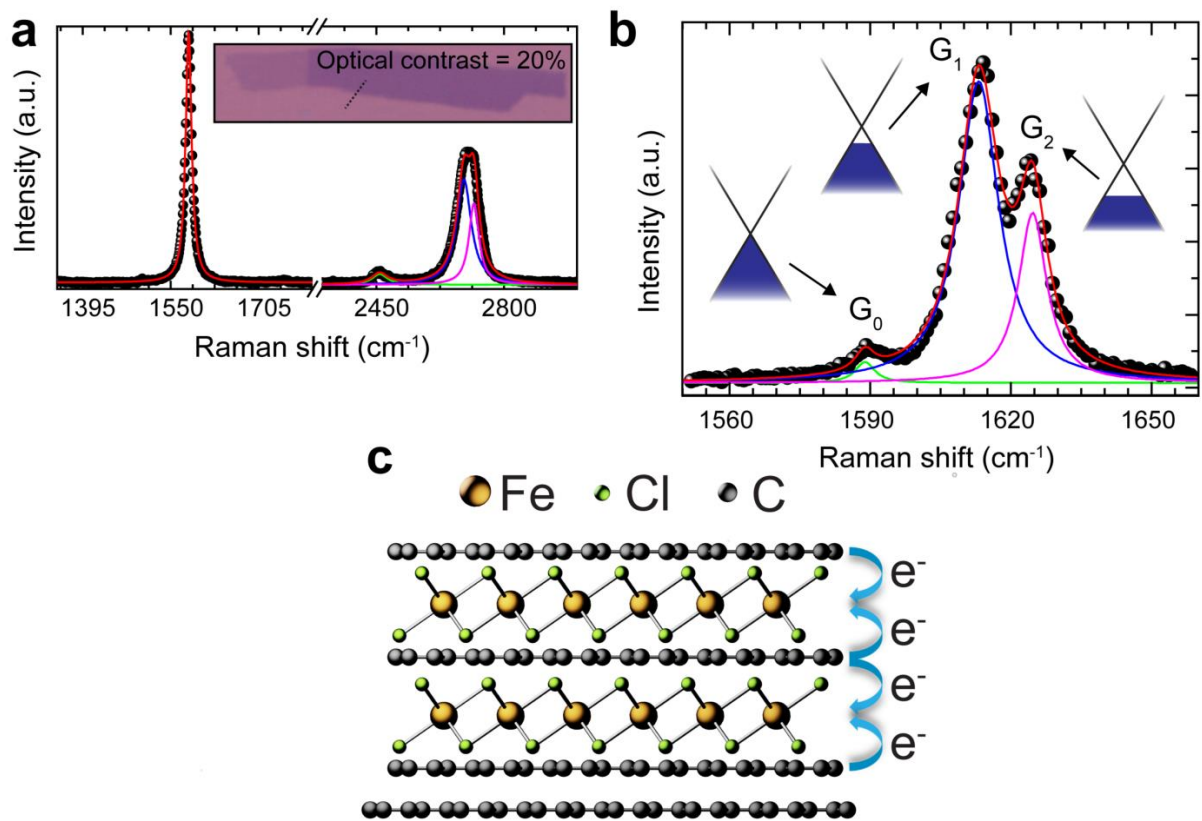

**Fig. S1. Inferred stacking order of four-layer FeCl$_3$-FLG.** (**a**) Raman spectrum of the same four-layer graphene flake before intercalation with FeCl$_3$. Inset, Image analysis of an optical micrograph shows a 20% contrast between the flake and Si/SiO$_2$ before intercalation. (**b**) Raman spectrum acquired after FeCl$_3$ intercalation, the levels of p-doping corresponding to the G$_0$, G$_1$ and G$_2$ peaks are illustrated. (**c**) Stacking order of the FeCl$_3$-FLG flake presented in figures 1-3 (main text).

Using a combination of optical microscopy and Raman spectroscopy it is possible to determine the stacking order of the FeCl$_3$-FLG. We consider the specific case of the flake discussed in the main text in figure 1a. This is a four-layer graphene as inferred from the optical contrast relative to the Si/SiO$_2$ substrate (20 %, under white light illumination) and the multi-peak structure of the Raman spectrum (figure S1a). Following FeCl$_3$ intercalation (*15*), we observe splitting of the G-band into three separate Lorentzian peaks (figure S1b). Each peak corresponds to a different level of charge carrier concentration due to a specific stage of intercalation (*15*). The G$_0$ peak at ~ 1585 cm$^{-1}$ corresponds to a pristine graphene layer, the G$_1$ peak at ~ 1610 cm$^{-1}$ to a graphene layer in contact with one layer (stage-2) and the G$_2$ peak at ~ 1625 cm$^{-1}$ to a graphene sheet sandwiched between two layers (stage-1). Hence, from the Raman spectrum we can identify the configuration reported in figure S1c. Here we have one graphene layer which remains isolated from FeCl$_3$. Two graphene layers are in contact with a single layer of intercalant and a fourth graphene layer at the centre of the structure is fully intercalated. It is highly improbable for FeCl$_3$ to remain on the top (or at the bottom) of the flake considering that any such layer would be directly exposed to all solvents used during subsequent device fabrication processes. Furthermore, the G$_1$ peak intensity is indicative of a larger presence of stage-2 intercalated states, relative to stage-1, as expected for the structure shown in figure S1c.

**S1.2 Exposure time and laser power effect**

In order to calibrate the laser-induced displacement of FeCl$_3$ with respect to the incident laser power and time, we performed a Raman spectroscopy study on two spots of a representative flake (shown in figure S2). The effect of exposing FeCl$_3$-FLG to laser powers of $0.15\ MW/cm^2$, $1.5\ MW/cm^2$, $4.1\ MW/cm^2$ and $15.3\ MW/cm^2$ is shown in figure S2a-c: it is evident that a change in G$_2$-peak height, indicative of a reduction in doping, only occurs upon exposure to a high-power light source. The dependence upon time was examined by irradiating a spot on the flake with a fixed power of $15.3\ MW/cm^2$ for $0$, $10$ and $600$ seconds (figure S2b-d). We observe that the doping modification happens very quickly, within the first 10 seconds, while a prolonged exposure causes no further effect (notably, the defect-related D-peak at $\sim 1350\ \text{cm}^{-1}$ does not emerge). Optical micrographs of the flake before and after laser exposure are shown in figure S2e, no visible modifications to FeCl$_3$-FLG are observed.

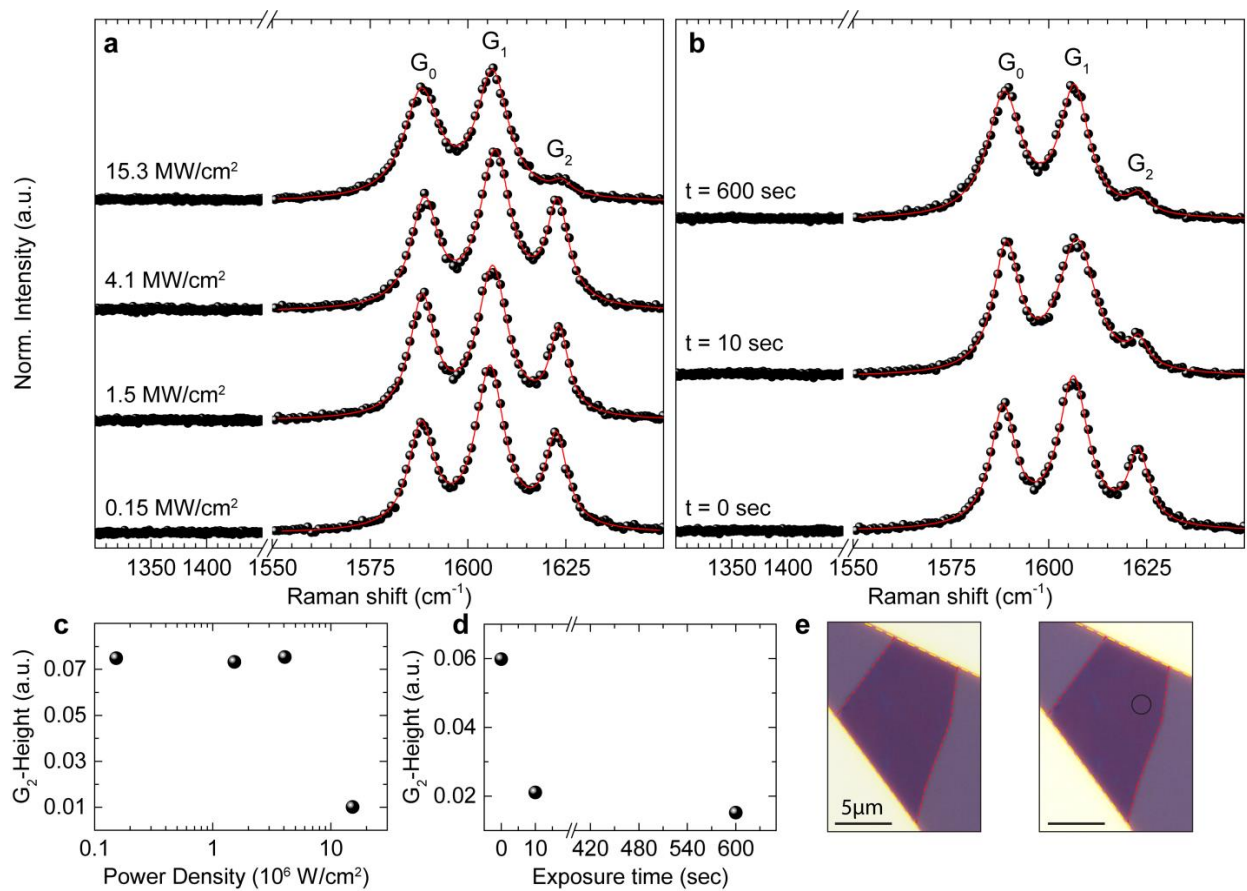

**Fig. S2. Calibration of laser-induced displacement of FeCl3.** (**a**) Raman spectra of FeCl$_3$-FLG acquired on the same location after irradiating with a $532\ nm$ laser light at different incident powers ($0.15\ MW/cm^2$, $1.5\ MW/cm^2$, $4.1\ MW/cm^2$ and $15.3\ MW/cm^2$) for 20 seconds. (**b**) Raman spectra of FeCl$_3$-FLG after irradiating with a power of $15.3\ MW/cm^2$ for 10 and 600 seconds compared with not-irradiated ($t = 0$ seconds). Each spectrum is acquired with the same laser at power of $0.15\ MW/cm^2$, red solid lines are Lorentzian fits. (**c**)-(**d**) Summary of the G$_2$-peak Height (normalized to the Si peak at $520\ cm^{-1}$) versus incident power and exposure time, as extrapolated from the fits in panels (**a**)-(**b**). (**e**) Optical micrograph of the examined FeCl$_3$-FLG flake before (right) and after (left) laser irradiation on the highlighted spot (black circle), no optical modifications are visible in the flake.

# S2 Supplementary photocurrent measurements

## S2.1 Bandwidth of FeCl₃-FLG photodetectors

In figure S3a we show the frequency-modulated photoresponse of the device presented in figure 3a in the main text. The $-3\,\text{dB}$ cut-off gives an operating bandwidth of $700 \pm 5\,\text{Hz}$, in good agreement with the rise and fall time measurements shown in figure S3b-c.

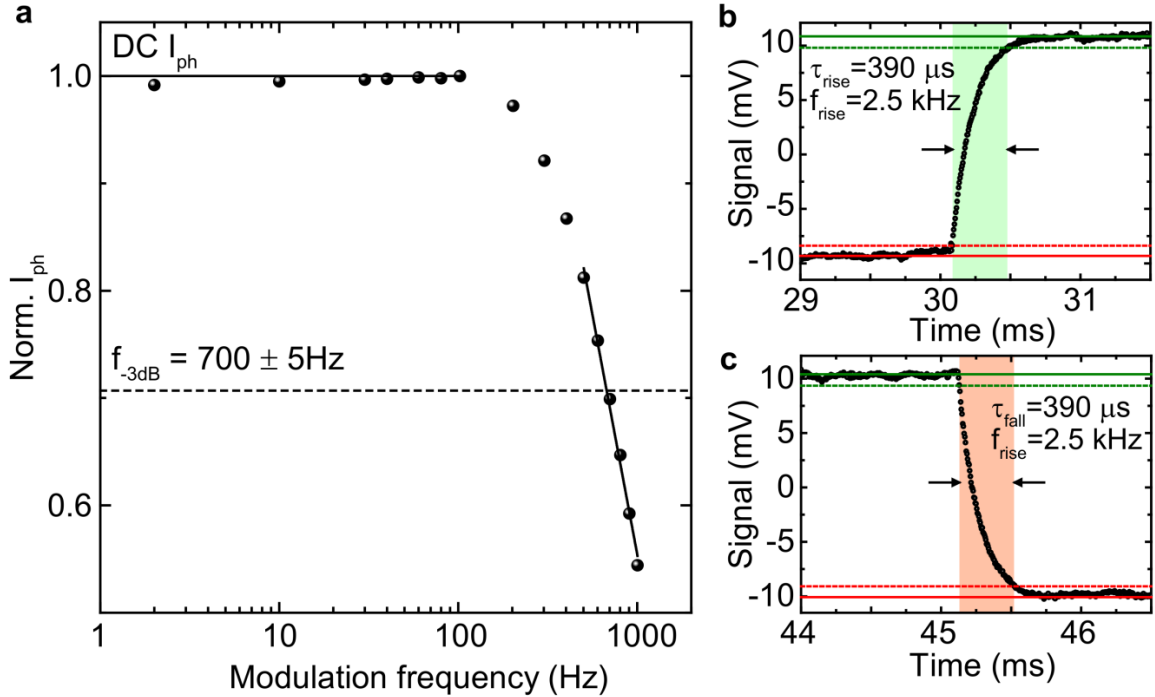

**Fig. S3. Bandwidth of a laser-written FeCl₃-FLG junction device.** (**a**) Frequency-modulated photoresponse of the device shown in figure 3a, main text: photocurrent is normalized to the DC value and the $-3\,dB$ cut-off is marked by the dashed line. (**b**) Rise and (**c**) fall time of the same device. Solid lines mark the steady state, dashed lines mark the $10\% - 90\%$ thresholds.

## S2.2 Noise equivalent power (NEP) measurement

RMS noise measurements were performed with a lock-in amplifier measuring the photocurrent directly with no current preamplifier in the circuit. The lock-in noise equivalent bandwidth (NEBW) was set to be $16.6\,\text{Hz}$, the modulation frequency was $689\,\text{Hz}$. Measured values are reported in figure S4 together with values of the photocurrent, as a function of incident laser power. The NEP is extrapolated to be $4\,\text{kW/cm}^2$.

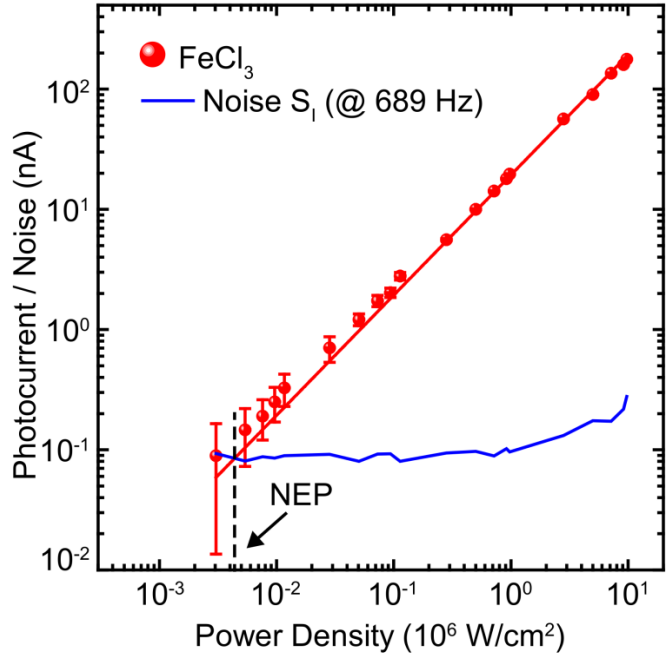

**Fig. S4. NEP of laser-written FeCl$_3$-FLG junction device.** Photoresponse as a function of laser power (red) together with the RMS noise measured during the same experiment (blue). The intersection marks the value of the NEP.

### S2.3 Comparison of the LDR of graphene photodetectors

In table S1, we show the saturation power density ($P_{\text{sat}}$) of graphene and functionalized graphene photodetectors reported in literature compared to the values measured in this work for FeCl$_3$-FLG junctions. Previous works have shown deviation from linear behaviour and saturation of photocurrent for power densities $< 57 \text{ kW/cm}^2$ in graphene (*9*) and $< 120 \text{ kW/cm}^2$ in functionalized graphene (*14*). In contrast, FeCl$_3$-FLG junctions show a saturation level $> 10^4 \text{ kW/cm}^2$, more than two orders of magnitude larger than other reports.

In the same table we report the linear dynamic range (LDR) in decibels ($\text{dB}$), calculated as:

$$LDR = 10 \times log_{10}\left(\frac{P_{sat}}{NEP}\right) [dB], \quad (S1)$$

where the Noise Equivalent Power (NEP) is defined as the power at which the signal to noise ratio (SNR) has a value of 1. The NEP can be measured directly or computed as:

$$NEP = \frac{S_I}{R} \left[\frac{W}{\sqrt{Hz}}\right], \quad (S2)$$

**Table S1. LDR of graphene and functionalized graphene devices.**

| Literature Reference | $P_{sat}$ [a] | NEP [b] | LDR [c] |
|---|---|---|---|
| Kim *et al.* (*17*) | $10^{-3}$ W/cm² | - | - |
| Liu *et al.* (*20*) | 1.27 W/cm² | 0.03 W/cm² | 15 dB [d] |
| Tielrooij *et al.* (*18*) | 23 kW/cm² | - | - |
| Mueller *et al.* (*3*) | 51 kW/cm² | 10 kW/cm² | 7.5 dB [e] |
| Graham *et al.* (*9*) | 57 kW/cm² | - | - |
| Patil *et al.* (*19*) | 14 kW/cm² | - | - |
| Wang *et al.* (*13*) | 120 kW/cm² | 3.3 kW/cm² | 15 dB [e] |
| This work (Graphene) | 45 kW/cm² | - | - |
| **This work (FeCl₃-FLG)** | $\mathbf{> 10^4}$ **kW/cm²** | **4 kW/cm²** | **44 dB** [d] |

[a] Power density at which saturation of photocurrent is observed; [b] Noise Equivalent Power; [c] Linear Dynamic Range; [d] Measured; [e] Estimated.

where $S_I$ is the RMS current noise (in $A/\sqrt{Hz}$) and $R$ is the responsivity of the photodetector (in $A/W$). We used equation S2 to calculate the NEP of different graphene-based photodetectors reported in literature (*9,13*). Assuming a graphene photodetector operating at the same frequency as our device ($689\,Hz$, see section S2.1), we can assume that the main source of noise will be the $1/f$ contribution (*38*). Using the results in references (*38*) and (*39*) we assume a spectral noise of $S_I = 1.0 \times 10^{-8}\,A/\sqrt{Hz}$. The NEP for reference (*20*) is taken from the measured values, the LDR agrees well with our estimation for the other references.

## S2.4 Photocurrent in pristine graphene

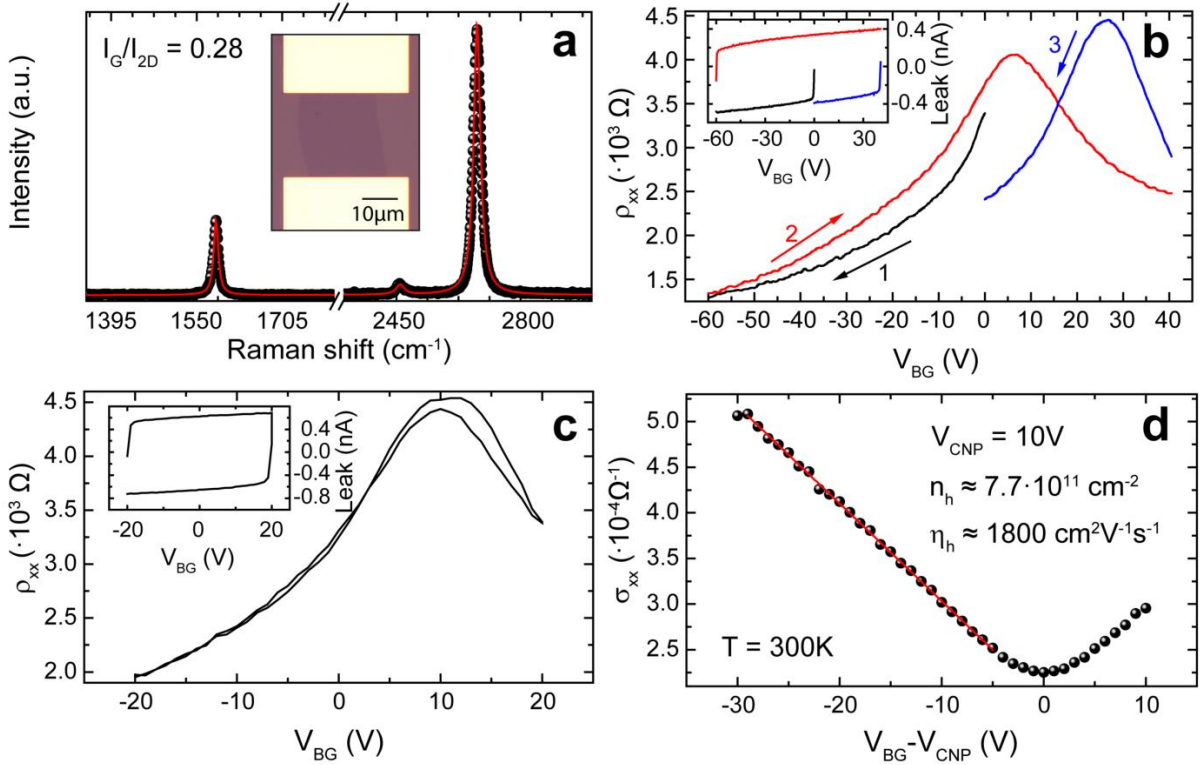

**Fig. S5. Characterization of supported pristine graphene devices.** (**a**) Raman spectrum of a monolayer graphene device. Inset: Optical micrograph of the same sample. (**b**) Longitudinal resistivity ($\rho_{xx}$) as a function of gate voltage ($V_{BG}$) for the device shown in panel (**a**) before ~20 hours in Acetone (60 °C) and rinsing for 1 hour in Isopropanol (60 °C). Numbers indicate the chronological sequence of gate voltage sweeps. (**c**) Gate sweeps of the same device after acetone-IPA treatment. Insets: gate leakage current as a function of gate voltage. (**d**) Conductivity ($\sigma_{xx}$) as a function of $V_{BG} - V_{CNP}$ with the extrapolated values for the charge concentration and mobility. All measurements are performed at room temperature in air.

Measurements shown in figure 3a-b of the main text (black dots) were performed on a pristine graphene device consisting of a monolayer flake mechanically-exfoliated onto p-doped Si with a 280 nm surface oxide. Cr/Au (5/50 nm respectively) electrodes were defined via electron-beam lithography using a PMMA resist followed by thermal evaporation of the metals and lift-off in Acetone. Figure S5a shows a representative Raman spectrum and optical micrograph of the resultant device. We fit both the G and 2D bands with a single Lorentzian, revealing a relative intensity of $I_G/I_{2D} = 0.28$. The optical contrast between the graphene and Si/SiO$_2$ substrate is 5% which, combined with a non-degenerate 2D band and $I_G/I_{2D} < 1$, signifies the presence of a graphene monolayer. Figure S5b shows the longitudinal resistivity ($\rho_{xx}$) as a function of back-gate voltage ($V_{BG}$) for the same device. From an initial gate sweep, the charge-neutrality point ($V_{CNP}$) is located around 0 V. However, a large hysteresis is observed during subsequent sweeps with a shift in $V_{CNP}$ by as much as 30 V. This behaviour is typical of graphene devices with a high degree of surface contamination (e.g. polymer residues from fabrication) measured in atmosphere. Surface contaminants induce charge-transfer which affects the capacitive gating effect (*40*). To minimise the effect of impurities on the surface of graphene, we soaked this device in warm Acetone (60 °C) for ~20 hours and then rinsed in warm Isopropanol for 1 hour. The gate response following this procedure is shown in Figure S5c where hysteresis effects are greatly reduced, resulting in a stable neutrality point at $V_{CNP} = 10V$. We extract the hole concentration ($n_h$) and field-effect mobility ($\eta_h$) of our device using the relationships $n_i = \epsilon V_{CNP}/et$ and $\eta_i = \sigma/en_i$, where $i$ indicates the polarity of charge carriers, $e$ is the electron charge, $t$ and $\epsilon$ are the thickness and absolute permittivity of respectively. The resulting values, shown in Figure S5d, are $n_h \approx 7.7 \cdot 10^{11}\ \text{cm}^{-2}$ and $\eta_h \approx 1800\ \text{cm}^2\text{V}^{-1}\text{s}^{-1}$. Having reduced the charge carrier concentration two orders of magnitude below that of FeCl$_3$-FLG layers, we performed the photo-current measurements shown in figure 3a (main text) at $V_{BG} = 0V$.

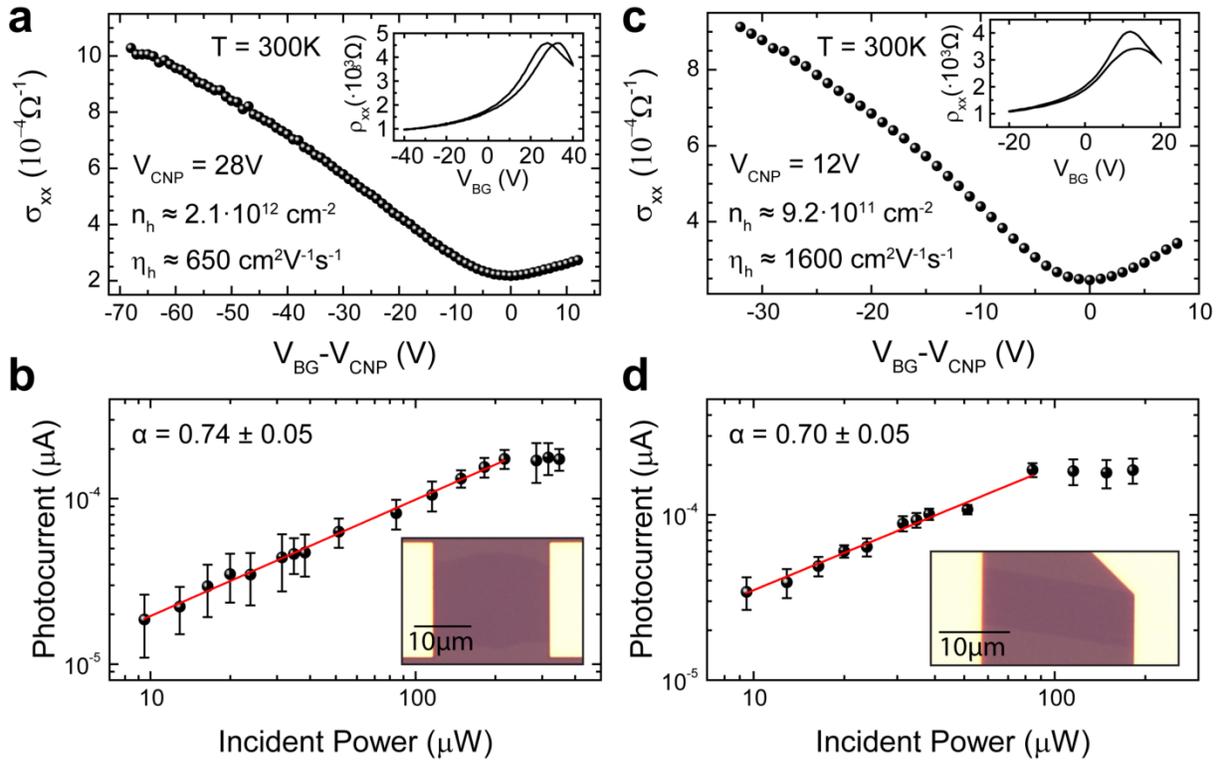

**Fig. S6. Additional measurements of photocurrent in supported pristine graphene devices.** (**a**) Conductivity ($\sigma_{xx}$) as function of gate voltage relative to the charge neutrality point device A. Inset, gate voltage dependence of resistivity. (**b**) Photo-current as function of laser incident power ($\lambda = 473\ nm$, $V_{BG} = 0\ V$) for device A. Inset, micrograph of device A. Equivalent measurements for device B are shown in panels (**c**) and (**d**). Measurements were taken in ambient conditions and at room temperature after prolonged soaking in warm acetone and Isopropanol (see figure S5).

Figure S6 shows the electrical and optoelectronic characterisation of two other pristine graphene devices (A and B respectively). Measurements were performed in ambient conditions after soaking each device in acetone for ~20 hours. Figure S6a and figure S6c show marginal differences in carrier concentration due to surface contamination. The power-dependence of the photocurrent ($I_{ph} \propto P_{opt}^\alpha$) measured in samples A (figure S6b) and B (figure S6d) were taken with a $\lambda = 473$ nm excitation laser, $V_{BG} = 0$ V and 10 mV applied between source and drain. Power-law exponents of $\alpha = 0.74 \pm 0.05$ and $\alpha = 0.70 \pm 0.05$ were extracted, both in agreement with dominant photothermoelectric effects observed in supported pristine graphene devices.

## S2.5 Photocurrent at p-p' junctions in FeCl$_3$-FLG

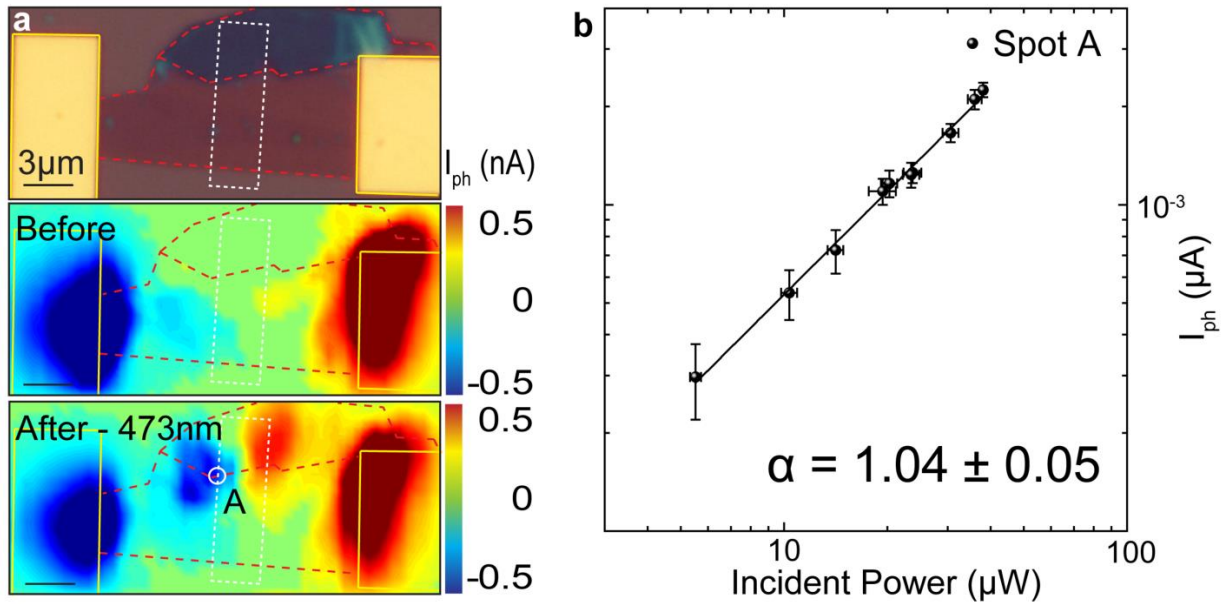

**Fig. S7. Photo-response at p-p' junctions in -FLG.** (**a**) Optical micrograph (top panel) and scanning photocurrent maps of a FeCl$_3$-FLG flake before (middle panel) and after (bottom panel) laser-induced de-intercalation. Superimposed lines indicate boundaries of the FeCl$_3$-FLG flake (red dashes), Au contacts (yellow) and de-intercalated area (white-dashes). Scale bars, $3 \mu m$. (**b**) Absolute photocurrent as a function of incident power measured at spot A (white circle in **a**) for $\lambda = 473\ nm$ excitation, a power exponent of $\alpha = 1.04 \pm 0.05$ is obtained from a fit to the experimental data (solid line).

Figure S7 presents photocurrent measurements at p-p' interfaces of FeCl$_3$-FLG in addition to those shown in the main text. All measurements were taken in short-circuit configuration with a two terminal device geometry. An optical micrograph image of the FeCl$_3$-FLG flake is shown in figure S7a where two distinct areas of different thickness are apparent. No substantial photocurrent is observed between these two regions either before or after laser-induced de-intercalation. After performing a raster scan with a 15.3 MW/cm$^2$ incident laser power ($\lambda = 532$ nm, 1 μm steps) over the region highlighted by the white dashed line, photocurrent was measured at the p-p' interfaces. The power dependence of this photocurrent (figure S7b) exhibits an exponent of $\alpha = 1.04 \pm 0.05$, similar to measurements shown in figure 3a-b of the main text.

## S3 Power dependence of the photothermoelectric and photovoltaic effects

### S3.1 Power dependence of the photothermoelectric (PTE) effect

The photothermoelectric (PTE) effect can exhibit a variety of power law ($I_{ph} \propto P_{opt}^{\alpha}$) exponents depending on the dominant cooling mechanism and the average temperature of hot carriers ($T_h$) relative to that of the lattice/environment ($T_l$). This is due to presence of a "bottleneck effect" whereby $T_h$ may remain above $T_l$ for photo-excited carriers in graphene due to the limited availability of pathways for heat dissipation. Initial coupling with high-energy optical phonon modes is exhausted for chemical potential ($\mu$) $< 200$ meV, leaving hot carriers to equilibrate through electron-electron scattering then gradually lose energy to the lattice (*41*). Heat dissipation is slow due to the small Fermi surface of graphene which limits energy losses through the momentum-conserving emission of an acoustic phonon ($\Delta E_{ac} < 2\hbar v_s k$ where $v_s \sim 2 \cdot 10^4$ ms$^{-1}$ is the acoustic phonon speed (*42*) and $k$ is the hot-carrier wavenumber) (*23*). The "supercollision" model (*9,24*) recognises that, in this situation, short-range scattering at sites of disorder allow a far larger transfer of energy and will be the dominant mechanism of carrier relaxation. The rate of heat loss ($H$) when supercollisions are dominant is given by:

$$H_{SC} = A(T_h^3 - T_l^3), \qquad A = \frac{9.62 g^2 D(\mu)^2 k_B^3}{\hbar k l}, \tag{S3}$$

where $g$ is the electron-phonon coupling frequency, $D(\mu)$ is the density of states and $l$ is the mean free path of hot carriers. Under continuous wave (CW) illumination, a steady-state is reached when the optical power imparted to hot carriers equals the power transferred to the lattice ($P_{in} = H_{SC}$). The electron temperature may be related to the photothermoelectric current using the Mott relation (*27*)

$$S = -\frac{\pi^2 k_B^2 T_h}{3e} \cdot \frac{1}{\sigma} \cdot \frac{\partial \sigma}{\partial \mu}, \tag{S4}$$

in conjunction with a general expression for the photothermoelectric voltage generated at the junction of two materials, $V_{PTE} = (S' - S)\Delta T$, to give (*5*):

$$I_{PTE} = \beta T_h (T_h - T_l), \tag{S5}$$

where

$$\beta = -\frac{\pi^2 k_b{}^2}{3e} \left[ \frac{1}{\sigma'} \cdot \frac{d\sigma'}{d\mu'} - \frac{1}{\sigma} \cdot \frac{\partial \sigma}{\partial \mu} \right]. \tag{S6}$$

Assuming that hot electrons stabilise at a temperature far above that of the lattice ($T_h >> T_l$), equation (S3) may be reduced to

$$T_h = (P_{in}/A)^{1/3}. \tag{S7}$$

Similarly, equation (S5) becomes

$$I_{PTE} = \beta T_h^2. \tag{S8}$$

Hence, the measured photocurrent should have a power dependence of

$$I_{PTE} = \beta \left(\frac{P_{in}}{A}\right)^{2/3}. \tag{S9}$$

This is the power exponent commonly measured in graphene photodetectors on Si/SiO$_2$ substrates.

In the case where the electron temperature is only marginally above that of the environment ($T_h - T_l << T_l$, as is common for measurements in CW illumination (*43*)) a Taylor expansion of equation (S3) about $T_h \approx T_l$ yields

$$P_{in} \approx 3AT_l{}^2 (T_h - T_l). \tag{S10}$$

Combining equation (S10) with equation (S5), we find an approximately linear dependence between photocurrent and power:

$$I_{PTE} = \frac{\beta P_{in}}{3AT_l} + \frac{\beta P_{in}{}^2}{9A^2 T_l{}^4} \approx \frac{\beta P_{in}}{3AT_l}. \tag{S11}$$

Table S2 compiles the power-law exponents obtained from equivalent calculations using models which base $P_{in}(T_h)$ purely upon acoustic phonon scattering (*9,24*). All models of the PTE effect predict an approximately linear dynamic range when $T_h \approx T_l$, this condition is most likely to be satisfied by measuring $I_{PTE}$ at room temperature and with low incident powers.

The relative contributions of acoustic phonon scattering ($H_{AP}$) and supercollisions ($H_{SC}$) to the rate of heat loss from photo-excited charge carriers is determined by the degree of disorder in the sample, the environmental temperature and the size of the Fermi surface (i.e. the level of doping) (*24*):

$$\frac{H_{SC}}{H_{AP}} = \frac{0.77}{kl}\frac{\left(T_h^2 + T_h T_l + T_l^2\right)}{T_{BG}^2}. \tag{S12}$$

Equation (S12) is valid when $k_B T \ll \epsilon_F$, where $\epsilon_F$ is the Fermi level. $T_{BG}$ is the Bloch-Grüneisen temperature of graphene (*26,44*) ($T_{BG} = \Delta E_{ac}/k_B$). The degree of disorder and doping will vary significantly between samples and therefore makes the wide variation in power dependence characteristics reported for graphene photodetectors understandable. In the case of FeCl$_3$-FLG, high levels of p-doping will significantly increase the Fermi surface thereby allowing larger energy losses via momentum-conserving acoustic phonon emission. As a result, hot carrier bottleneck effects will be less prominent and the contribution of defect-assisted scattering towards photocurrent in FeCl$_3$-FLG is likely to be small compared to interfaces in graphene photodetectors with low levels of doping.

**Table S2. Summary of power-law exponents possible for photocurrent originating from the photothermoelectric effect.**

| PTE Model | $P_{in}(T_e)$ | $\alpha$ ($T_e \gg T_l$) | $\alpha$ ($T_e - T_l \ll T_l$) |
|---|---|---|---|
| **Supercollision** | $A(T_e^3 - T_l^3)$ | 2/3 | $\approx 1$ |
| **Acoustic** ($k_B T \gg \epsilon_F$) | $A' T_e^4 (T_e - T_l)$ | 2/5 | $\approx 1$ |
| **Acoustic** ($k_B T \ll \epsilon_F$) | $A''(T_e - T_l)$ | 2 | $\approx 1$ |

**S3.2 Power dependence of the photovoltaic (PV) effect**

The photovoltaic effect describes the separation of an electron-hole pair by an in-built electric field. In the low-power regime where the photocarrier lifetime, $\tau_c$, is independent of the photo-generation rate, $r_g$, photocurrent may be shown to have a linear dependence upon incident power, with $I_{PVE} \propto P_{opt}$. For a photodetection layer the steady-state photo-generation rate of carriers is given by (*25*):

$$r_g = \frac{\chi \Phi_{ph}}{A_{ph}D}, \tag{S13}$$

where $\chi$ is the quantum efficiency of the absorption process, $\Phi_{ph}$ is the incident photon flux, $A_{ph}$ is the illuminated area and $D$ is the thickness of the layer. The recombination rate of excess carriers depends on the minority carrier lifetime $\tau_c$ via:

$$r_r = \frac{n}{\tau_c} = \frac{p}{\tau_c}, \tag{S14}$$

where $n$ and $p$ are the excess carriers populations. Therefore, in equilibrium the generation rate must equal the recombination rate and the photocarrier density is:

$$n = p = r_g \tau_c = \frac{\chi \Phi_{ph} \tau_c}{A_{ph}D}. \tag{S15}$$

Given a potential difference $V$, between the sides of the layers, a photoinduced current $I_{PVE}$ can be measured

$$I_{PVE} = \frac{WD}{L}\sigma V = \frac{WD}{L} r_g \tau_c e (\eta_e + \eta_h) V, \tag{S16}$$

where $\sigma = ne\eta_e + pe\eta_h$ is the electrical conductivity, $\eta_h$ and $\eta_e$ are the hole and electron mobilities, $W$ and $L$ are the width and the length of the channel and $e$ is the electron charge. Combining equation (S15) and equation (S16) and noting that $\Phi_{ph} = P_{opt}/h\nu$, where $P_{opt}$ is the incident optical power, $h$ is Plank's constant and $\nu$ is the frequency of the incident light, we arrive to the final expression:

$$I_{PVE} = \eta \frac{P_{opt}}{h\nu} \frac{e(\eta_e + \eta_h)V\tau_c W}{A_{ph}L}. \tag{S17}$$

Hence we can define the photoconductive gain $G$ as the ratio of the rate of flow of electrons per second from the device to the rate of generation of e-h pairs within the device

$$G = \frac{I_{PVE}}{e} \frac{1}{r_g WDL} = \frac{\tau_c (\eta_e + \eta_h) V}{L^2}. \tag{S18}$$

Equation (S17) shows the relation $I_{PVE} \propto (P_{opt})^\alpha$ with $\alpha = 1$.

# S4 Estimation of chemical potential and conductivity for decoupled graphene layers

## S4.1 Estimation of chemical potential

In order to explain the physical mechanisms responsible of the measured photoresponse at p-p' junctions in FeCl$_3$-FLG, it is necessary to estimate the chemical potential of an intercalated flake before and after laser irradiation. Previous studies have shown through Raman spectroscopy (*45*) and magneto-transport measurements (*14*) that highly intercalated samples of FeCl$_3$-FLG may be considered as parallel stacks of electrically isolated monolayers. Using the density of states for monolayer graphene, we define the chemical potential ($\mu$) of each decoupled graphene sheet as $\mu = \hbar v_F \sqrt{\pi n}$ where $v_F \approx 10^6$ ms$^{-1}$ is the Fermi velocity and $n$ is the density of holes. Note that we equate the chemical potential of our system with the Fermi level, as $k_B T \ll \epsilon_F$ for all situations relevant to our discussion. Following the reasoning given in section S1.1, the flake shown in figure S8a is four-layer graphene the bottom two sheets remain electrically coupled as a bilayer (figure S8b). Going from top to bottom, we now refer to the decoupled graphene systems as $A$, $B$ and $C$ (figure S8b). Dash terms (e.g. $\sigma_A'$) represent a material's properties after laser-induced de-intercalation.

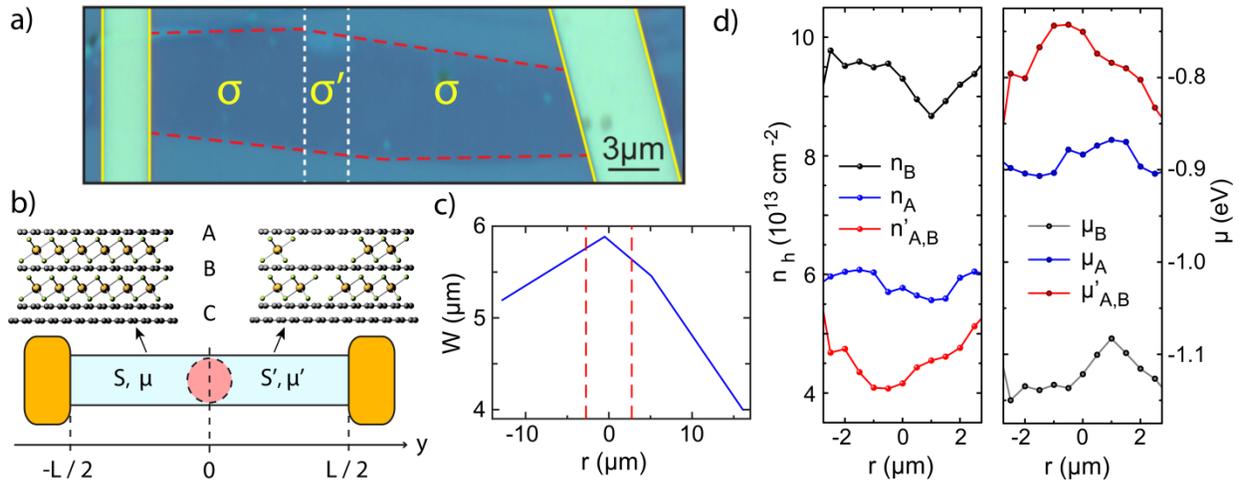

**Fig. S8. Calculation of the carrier concentration and chemical potential at p-p' interfaces of FeCl$_3$-FLG.** (**a**) Micrograph picture of a four-layer FeCl$_3$-FLG flake with a p-p'-p junction patterned by $\lambda = 532\ nm$ laser irradiation (main text). Superimposed lines represent boundaries of the flake (red), contacts (yellow) and the de-intercalated area (white). (**b**) Schematic of a p-p' interface located at the centre of a long, narrow FeCl$_3$-FLG channel. The degree of intercalation, inferred from Raman spectroscopy measurements, is illustrated for each region with the three decoupled systems labelled $A$, $B$ and $C$. (**c**) Width of the flake shown in (a) as a function of distance along the channel length. Red dashes mark the boundaries of the p' region. (**d**) Concentration of charge carriers in decoupled graphene layers inferred from the position of the G$_1$ and G$_2$ Raman peak positions shown in figure 1c (main text). The chemical potential is then calculated using the density of states for monolayer graphene.

The chemical potential of the bilayer system $C$ will not be affected as radically as the monolayers $A$ or $B$ when in proximity to 1 layer of FeCl3, we therefore focus our discussion on the upper two layers of the flake. In figure S8d, the model of Lazzeri *et al.* (*21*) is used to convert from the positions of G$_1$ and G$_2$ Raman peaks to the carrier concentration in each layer before and after laser writing ($n'_A \approx n'_B$ after irradiation). Taking a linear band approximation, the respective chemical potentials are plotted in figure S8d, giving average values of $\mu_A = (-0.88 \pm 0.02)eV$, $\mu_B = (-1.12 \pm 0.2)eV$ and $\mu'_{A,B} = (-0.76 \pm 0.02)eV$. Marginally smaller shifts in Fermi level have been measured in intercalated graphene grown by chemical vapour deposition (*15*), but our estimated values agree well with those previously reported in DFT calculations (*28*) and Raman spectroscopy measurements (*14,45*) of exfoliated flakes.

### S4.2 Estimation of conductivity

Two terminal resistance measurements of the FeCl3-FLG flake in figure S8a were taken before and after laser patterning using a lock-in amplifier in constant current configuration. Through image analysis, we calculate the change in channel width along the entire flake (figure S8c) and relate it to the conductivity, $\sigma_{tot}$, of the two different regions:

$$R_{SD} = \frac{1}{\sigma_{tot}} \int_{-L/2}^{L/2} \frac{1}{W(y)} dy, \tag{S19}$$

$$R'_{SD} = \frac{1}{\sigma_{tot}} \left[ \int_{-L/2}^{y_1} W(y)^{-1} dy + \int_{y_2}^{L/2} W(y)^{-1} dy \right] + \frac{1}{\sigma'_{tot}} \int_{y_1}^{y_2} W(y)^{-1} dy, \tag{S20}$$

where $y_1$ and $y_2$ denote the boundaries of the irradiated p' area. Through equations (S19) and (S20) we find $\sigma_{tot} \approx 27$ mS and $\sigma'_{tot} \approx 10$ mS, slightly below the maximum conductivity of fully intercalated four-layer flakes.[14] Approximating $n_{tot} \approx 2n_A + n_B$ and $n'_{tot} \approx 3n'_{A,B}$, the average hole mobility is taken to be $\langle \eta \rangle = 650$ cm$^2$V$^{-1}$s$^{-1}$. Lastly, we attain conductivity values for the individual systems A and B using

$$\sigma(\mu) = \frac{e\eta\mu^2}{\pi\hbar^2 v_F^2} + \sigma_{min}, \tag{S21}$$

where $\sigma_{min} \sim 4e^2/h$ (*46*). This may also be written in the form

$$\sigma(\mu) = \sigma_{min}\left(1 + \frac{\mu^2}{\Lambda^2}\right) \quad , \quad \Lambda \approx 140 meV. \tag{S22}$$

We find $\sigma_A = 6.0$ mS, $\sigma_B = 9.6$ mS and $\sigma'_{A,B} = 4.5$ mS.

# S5 Physical explanation for a purely photovoltaic response

Here, we estimate the relative magnitudes of photocurrent produced by the photovoltaic and photothermoelectric effects at a p-p' junction of FeCl$_3$-FLG. We consider a single junction located in the middle of an FeCl$_3$-FLG channel (figure S8b) in order to simplify our explanation of the underlying photoresponse mechanisms and demonstrate that the suppression of thermoelectric currents in our devices is not simply due to the proximity of two junctions with opposing polarity. Following a similar method to Song *et al.* (*5*), the total photocurrent produced when the interface is illuminated, under short circuit conditions, is taken to be a summation of photovoltaic and thermoelectric contributions:

$$I_{PH} = \frac{1}{RW} \int_0^W \int_{-L/2}^{L/2} \left[ S(x,y) \nabla T(x,y) - \sigma(x,y)^{-1} e\eta n_{ph}(x,y) \nabla U(x,y) \right] dy dx. \quad (S23)$$

The first term of the integral represents thermoelectric currents produced by a temperature gradient $\nabla T(x,y)$ in a material with a spatially varying Seebeck coefficient $S(x,y)$. The second term describes the photovoltaic response produced when a density $n_{ph}(x,y)$ of carriers are generated in a material and then displaced by an in-built potential gradient $\nabla U(x,y)$.

## S5.1 Photothermoelectric Effect (PTE)

Approximating $S(y)$ as a step change at the p-p' junction and substituting equation (S22) into equation (S4), we re-write the PTE current in terms of the electrical properties of the regions either side of the p-p' interface (figure S8a):

$$I_{PTE} = \frac{2\pi^2 k_B^2 T_h}{3eR} \cdot \frac{\Delta T}{\mu\mu'} \cdot \left[ \mu' \left(1 - \frac{\sigma_{min}}{\sigma}\right) - \mu \left(1 - \frac{\sigma_{min}}{\sigma'}\right) \right]. \quad (S24)$$

The difference in steady state temperature between the lattice and hot carriers ($\Delta T = T_h - T_l$) is a difficult quantity to measure, requiring picosecond resolution of photocurrent transients in low temperature environments (*9*) which are beyond the scope of our experimental apparatus. Alternative methods which approximate values of $\Delta T$ using equation (S24) rely on the assumption that any measured photovoltage is produced solely by thermoelectric currents (*7*). This inference cannot be made for FeCl$_3$-FLG interfaces; extremely high carrier densities (up to $3 \times 10^{14}$ cm$^{-2}$ per layer) efficiently screen electrostatic gating potentials and prohibit experimental methods which are typically used to verify the "six fold pattern" signature of the PTE effect (*5,7,9,43*). Instead, we use a solution obtained for the one-dimensional heat equation of our system, where the photocurrent density created at the p-p' junction is assumed to be a delta function with respect to the laser spot size (*5*):

$$\Delta T = \frac{\alpha \epsilon_0 l_0 N_{ph}}{\frac{\kappa}{\zeta} \coth\left(\frac{L}{2\zeta}\right) + \frac{\kappa'}{\zeta'} \coth\left(\frac{L}{2\zeta'}\right) + \frac{T_0}{RW}(S'-S)^2}. \qquad (S25)$$

Here, $\alpha$ is the fraction of an absorbed photon's energy ($\epsilon_0$) which is retained by the hot electron system once electron-electron interactions and coupling with optical phonons have been exhausted. $l_0$ is the laser spot diameter and $N_{ph}$ represents the flux of absorbed photons at the centre of the p-p' junction averaged over the channel width. $\kappa$ and $\zeta$ are the thermal conductivity and average cooling length of hot electrons respectively. Provided $k_B T \leq (\mu, \mu', \Delta)$, the third term of the denominator in equation (S25) is negligible. The cooling length of each graphene layer is dependent upon its electrical conductivity, density of states ($D(\mu)$) and the hot carrier cooling rate ($\gamma$) (7):

$$\zeta = \sqrt{\frac{\sigma}{\gamma e^2 D(\mu)}}. \qquad (S26)$$

Naturally, $\gamma$ is dependent upon the prevailing hot electron scattering mechanism. For graphene layers where $n \geq 10^{13}$ cm$^{-2}$, the Bloch-Grüneisen temperature reaches hundreds of Kelvin and hot electrons may completely equilibrate with the lattice via just a single acoustic phonon interaction under CW illumination (26). Disorder-mediated scattering is therefore not relevant in our devices. This can be shown by substituting equation (S21) into equation (S12) using the relation for the mean free path of a non-degenerate two-dimensional electron gas, $l = \sigma \hbar \pi / k e^2$, to estimate the relative magnitudes of power loss via supercollisions and momentum-conserving scattering events in FeCl₃-FLG. For $T_h - T_l \ll T_l$, we find supercollisions to make up as little as 3% (11%) of the total heat loss from hot electrons before (after) laser-induced de-intercalation. The scattering rate can therefore be approximated by considering just single acoustic phonon processes (23) as:

$$\gamma = \frac{3D^2 \mu^3}{4\pi^2 \hbar^3 \rho_m v_F^4 k_B T_{el}}, \qquad (S27)$$

where $D \sim 20$ eV is the typical screened deformation potential on Si/SiO₂ substrates (26) and $\rho_m = 7.6 \cdot 10^{-7}$ kg m$^{-2}$ is the mass density of monolayer graphene. Due to the doping induced by FeCl₃ intercalation, the cooling rate of momentum-conserving acoustic phonon coupling dramatically increases from $\gamma \sim 10^{-9}$ s$^{-1}$ at $\mu = 100$ meV to $\gamma_A = 6 \cdot 10^{11}$ s$^{-1}$, $\gamma_B = 1 \cdot 10^{12}$ s$^{-1}$ and $\gamma'_{A,B} = 4 \cdot 10^{11}$ s$^{-1}$. This is in agreement with the picosecond relaxation time-scales of FeCl₃-FLG measured via pump-probe spectroscopy (28). Hence, we use equation (S26) to calculate cooling lengths of $\zeta_A =$ 220 nm, $\zeta_B = 170$ nm and $\zeta'_{A,B} = 260$ nm. Given that $\zeta \ll L/2$ for all of our devices, equation (S25) simplifies to:

$$\Delta T \approx \alpha \epsilon_0 l_0 N_{ph} \left(\frac{\kappa}{\zeta} + \frac{\kappa'}{\zeta'}\right)^{-1}. \tag{S28}$$

Substituting equation (S28) into equation (S24) and employing the Wiedemann-Franz relation (*47*), we arrive at a full expression for the photothermoelectric current produced at a p-p' junction in FeCl$_3$-FLG:

$$I_{PTE} = \frac{2eqk_B T_{el} l_0 N_{ph}}{\mu\mu' R} \cdot \left[\mu'\left(1 - \frac{\sigma_{min}}{\sigma}\right) - \mu_1\left(1 - \frac{\sigma_{min}}{\sigma'}\right)\right] \cdot \left[\frac{\sigma}{\zeta} + \frac{\sigma'}{\zeta'}\right]^{-1}, \tag{S29}$$

where $q \sim \alpha\epsilon_0/k_B T_{el}$ is the internal quantum efficiency.

### S5.2 Photovoltaic Effect (PVE)

From equation (S23), the photovoltaic contribution to the photocurrent is

$$I_{PVE} = -\frac{1}{RW}\int_0^W \int_{-\frac{L}{2}}^{\frac{L}{2}} \sigma(x,y)^{-1} e\eta n_{ph}(x,y) \nabla U(x,y) dy dx. \tag{S30}$$

Taking all values as averages over the channel width, $e\nabla U(y) = \nabla\mu(y)$ and using equation (S22), equation (S30) may be simplified as:

$$I_{PVE} = -\frac{\eta\, n_{ph\,(y=0)}}{\sigma_{min} R}\int_{-\frac{L}{2}}^{\frac{L}{2}} \nabla\mu(y) \cdot \left(1 + \frac{\mu(y)^2}{\Lambda^2}\right)^{-1} dy. \tag{S31}$$

By changing variables, we find a complete expression for the photovoltaic contribution to photocurrent:

$$I_{PVE} = \frac{qN_{ph}\eta\Lambda}{\sigma_{min}R\langle\gamma\rangle} \cdot \left[\tan^{-1}\left(\frac{\mu}{\Lambda}\right) - \tan^{-1}\left(\frac{\mu'}{\Lambda}\right)\right]. \tag{S32}$$

Here, we have approximated the steady state density of photogenerated carriers at the p-p' junction as $n_{ph\,(y=0)} \approx qN_{ph}/2\langle\gamma\rangle$ where $\langle\gamma\rangle$ is the average cooling rate of hot carriers over both sides of the p-p' junction and the average lifetime of a photogenerated carrier is $\tau \sim \langle\gamma\rangle^{-1}$.

### S5.3 Relative magnitudes of the PTE and PVE

Dividing equation (S29) by equation (S32), the relative magnitudes of photothermoelectric and photovoltaic currents at FeCl$_3$-FLG p-p' junctions may be calculated:

$$\frac{I_{PTE}}{I_{PVE}} = \frac{2ek_B T_{el} l_0 \langle \gamma \rangle}{\eta \Lambda} \cdot \frac{\left[\mu'\left(1 - \frac{\sigma_{min}}{\sigma}\right) - \mu_1\left(1 - \frac{\sigma_{min}}{\sigma'}\right)\right]}{\mu\mu'\left(\frac{\sigma}{\zeta} + \frac{\sigma'}{\zeta'}\right)\left[tan^{-1}\left(\frac{\mu}{\Lambda}\right) - tan^{-1}\left(\frac{\mu'}{\Lambda}\right)\right]}. \tag{S33}$$

For both decoupled systems $A$ and $B$ we calculate $I_{PTE}/I_{PVE} \approx -0.06$, hot carrier dynamics therefore make a negligible contribution to the total photocurrent generated at FeCl$_3$-FLG p-p' junctions and act in the opposite direction to currents produced by the photovoltaic effect.

### S5.4 Direction of photocurrent at p-p' junctions in FeCl$_3$-FLG

Based upon previously reported theoretical models (*5*) for graphene-based photodetectors with split electrostatic gates and equation (S33), photothermoelectric currents in graphene will travel in the opposite direction to photovoltaic currents at p-p' and n-n' junctions. This is due to the additional polarity change which PTE currents undergo which is often illustrated by the "six fold pattern" of photocurrent measured at dual-gated interfaces (*7,9,43*). Taking advantage of this asymmetry, we examine the direction of the photocurrent measured at p-p'-p junctions in order to further confirm that the PVE is indeed dominant in laser-written FeCl$_3$-FLG photodetectors.

Figure S9a shows a scanning photocurrent map taken from the main text of a laser-irradiated FeCl$_3$-FLG flake with a p-p'-p junction. This measurement was performed with source and drain electrodes grounded and a current amplifier (DL Instruments, Model 1211) connected in series with the left electrode which sends an output voltage signal, $V_{OUT}$, to a lock-in amplifier. Calibrating this measurement circuit with a known DC voltage input, we find that positive (red) photocurrent in figure S9a signifies the drift of holes to the right electrode and electrons to the left. If hot carrier dynamics are suppressed, photocurrent at laser-written interfaces of FeCl$_3$-FLG will flow in the direction illustrated in figure S9b, where charges drift with respect to the local potential gradient. However, if PTE effects dominate the measured photoresponse then the configuration illustrated in figure S9c is expected. Comparing figure S9a with each of these two scenarios, it is clear that the photocurrent measured at p-p'-p interfaces of FeCl$_3$-FLG is predominantly produced by the photovoltaic effect.

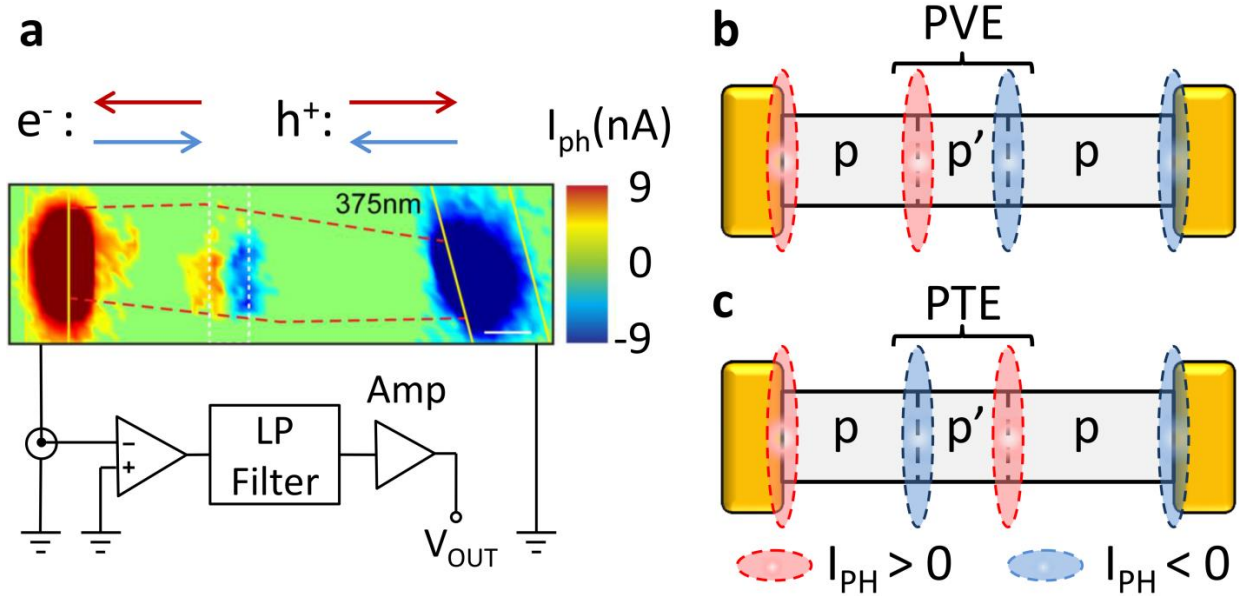

**Fig. S9. Direction of photocurrent at p-p' junctions of FeCl$_3$ -FLG.** (**a**) Scanning photocurrent map of a p-p'-p junction in FeCl$_3$-FLG taken from figure 2 (main text). Measurements were taken in short-circuit configuration with an inverting current amplifier connected to the left electrode. Positive (red) signals indicate holes drifting to the right. Two schematics of the same device illustrate the predicted direction of photocurrent local to the junctions assuming that either (**b**) the photovoltaic (PV) or (**c**) the photothermoelectric (PTE) effects is the dominant mechanism of photoresponse.

## S6 Correction of responsivity spectra for substrate reflections

The presence of a reflecting Si/SiO$_2$ substrate will affect the measured spectral responsivity of our FeCl$_3$-FLG photodetectors. As shown in figure 3c (main text), we have performed a correction which accounts for these reflections in order to examine the intrinsic spectral response of the laser-written p-p' junctions. Figure S10a illustrates the model used for this correction which consists of an incident photon flux ($\Phi_0$) partially absorbed by an FeCl$_3$-FLG flake of transmittance $T$ and a transmitted remaining flux, $\Phi_t = T\Phi_0$. A portion of this transmitted flux ($\Phi_r = \Phi_t R$, where $R$ is the reflectance of Si/SiO$_2$) will be reflected by the substrate and absorbed/transmitted by the FeCl$_3$-FLG, leaving a flux $\Phi_{t'} = T\Phi_r$ reflected into the environment. We neglect further reflections due to the high transmittance of FeCl$_3$-FLG and define the spectral responsivity as $\Re(\lambda) = I_{ph}/\epsilon_0 \Phi$. Hence, the photon flux incident on a supported FeCl$_3$-FLG detector is effectively ($\Phi_0 + \Phi_r$) and the ratio between the measured ($\Re$) and intrinsic ($\Re_0$) responsivity may be evaluated using just $T$ and $R$:

$$\frac{\mathcal{R}_0}{\mathcal{R}} = \frac{\Phi_0}{\Phi_0 + \Phi_r} = \frac{1}{1+TR}. \tag{S34}$$

Figure S10b shows the transmittance of a four-layer FeCl$_3$-FLG sample reproduced with permission from reference (*14*) and the reflectivity of our Si/SiO$_2$ substrate measured in the range $420 - 700$ nm. A simulation of the substrate reflectivity using TFCalc software (Software Spectra, Inc.) shows excellent agreement with the experimental data, we therefore extrapolate the reflection coefficient from the simulated curve down to $\lambda = 375$ nm where no experimental data points are available. In the same way, we extrapolate the absorption coefficient of FeCl$_3$-FLG for the same wavelength range. The extrapolated data and the computed correction factors used in figure 3c (main text) are presented in table S3.

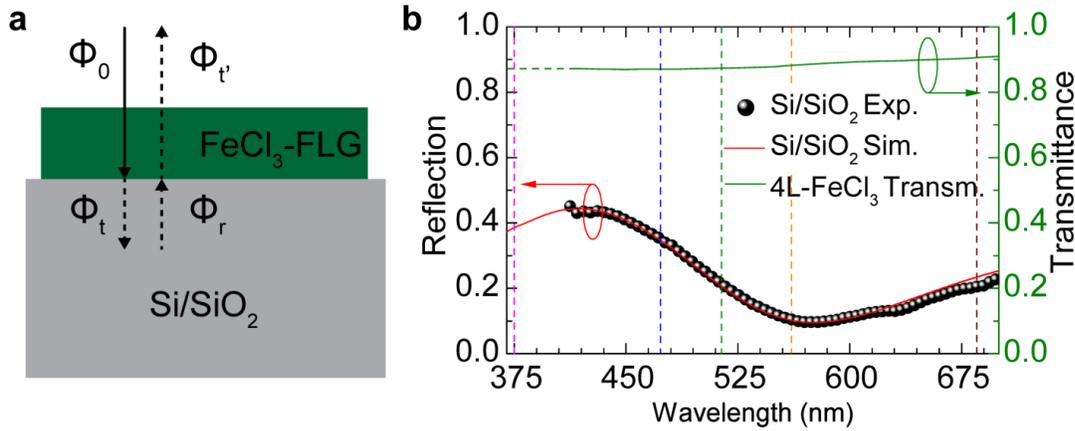

**Fig. S10. Correction of spectral responsivity for substrate reflections**. (**a**) Concept of substrate reflection correction of responsivity: solid arrow is the incoming light ($\Phi_0$), dotted lines represent the transmitted light through the FeCl$_3$-FLG ($\Phi_t$) and the reflected part by the Si/SiO$_2$ interface ($\Phi_r$). (**b**) Reflectivity of Silicon substrate with $290$ nm of SiO$_2$ on top: experimental values (black dots) in the region $420 - 700$ nm and computed curve (solid red line) between $370 - 700$ nm; the green line represents the transmittance of 4-layer FeCl$_3$-FLG (reproduced with permission from reference (*14*)) where we extrapolated the value for the UV-A region (dotted green line). Vertical dotted lines represent the laser wavelengths used in this work.

**Table S3. Corrections to responsivity for the laser wavelengths used in this work.**

| $\lambda$ (nm) | $T$ | $R$ | $\mathcal{R}_0/\mathcal{R}$ |
|---|---|---|---|
| 375 | 0.872 | 0.385 | 0.749 |
| 473 | 0.870 | 0.355 | 0.764 |
| 514 | 0.874 | 0.207 | 0.847 |
| 561 | 0.883 | 0.102 | 0.917 |
| 685 | 0.906 | 0.234 | 0.825 |